\documentclass[12pt]{article}
\usepackage{latexsym,graphicx,multirow}
\usepackage{amssymb}
\usepackage{amsmath}
\usepackage{amscd}
\usepackage{amsthm}
\usepackage{float}
\usepackage[left=2cm,top=2.5cm,right=2.5cm,bottom=1.5cm]{geometry}
\usepackage{graphicx}
\usepackage{epstopdf}
\usepackage{graphicx,epstopdf}
\usepackage{hyperref}
\DeclareGraphicsExtensions{.eps}
\usepackage{epstopdf} 
\usepackage{pdflscape}
\usepackage{cite}
\begin{document}
	\begin{center}
		\large{\bf Interacting R$\acute{e}$nyi holographic dark energy with parametrization on the interaction term} \\
		\vspace{10mm}
		\normalsize{Umesh Kumar Sharma$^1$, Vipin Chandra Dubey$^2$}\\
		\vspace{5mm}
		\normalsize{$^{1,2}$Department of Mathematics, Institute of Applied Sciences and Humanities, GLA University\\
			Mathura-281 406, Uttar Pradesh, India}\\
		\vspace{2mm}
		$^1$E-mail: sharma.umesh@gla.ac.in.\\
		$^2$E-mail: vipin.dubey@gla.ac.in\\
		
		\vspace{2mm}
		\vspace{5mm}
		\vspace{10mm}

	\end{center}
	
	\begin{abstract}
		
		In the present work, we study the R$\acute{e}$nyi holographic dark energy model (RHDE) in a flat FRW Universe where the infrared cut-off is taken care by the Hubble horizon and also by taking three different parametrizations of the interaction term between the dark matter and the dark energy. Analysing graphically, the behaviour of some cosmological parameters in particular deceleration parameter, equation of state (EoS) parameter, energy density parameter and squared speed of sound, in the process of the cosmic evolution, is found to be leading towards the late-time accelerated expansion of  the RHDE model.
	\end{abstract}

	

	\section{Introduction}
	Our Universe is undergoing accelerated expansion which is marked by various cosmological observations like type-Ia supernova \cite{ref1,ref2,ref3,ref4}, the large-scale structure\cite{ref5,ref6,ref7,ref8}, cosmic microwave background (CMB) anisotropies \cite{ref9,ref10,ref11}. For explaining this accelerated expansion of the cosmos the concept of dark energy (DE) was incorporated which is an extraordinary component with negative pressure \cite{ref12,ref13}. The late-time acceleration of the Universe can be explained by two  categories models. First one is dynamical dark energy models in which we change the matter part of the Einstein field equation. Amongst a lot of theories and models the cosmological constant model is the simplest model, initially proposed by  Einstein\cite{ref14a,ref14b,ref14c,ref14d,ref14e}, which suggests that the equation of the state parameter (EoS) $\omega = - 1$ and  the cosmological constant is the most basic candidate for dark energy, and it is consistent with observations, besides the fine-tuning and coincidence problem \cite{ref14b,ref14f,ref14g}.
	To get relief from such problems, many dynamical dark energy (DE) models are given, for example, $k$-essence \cite{ref14h}, quintessence \cite{ref14i,ref14j}, Chaplygin gas \cite{ref14k}, phantom \cite{ref14l}, tachyon \cite{ref14m}, holographic dark energy (HDE) \cite{ref14n} and new agegraphic dark energy (NADE)\cite{ref14q}.  Secondly, by modified gravity theories such as $f(R)$ theory \cite{ref14t}, $f(T)$ theory \cite{ref14u}, Ho$\check{r}$ava-Lifshitz gravity \cite{ref14v,ref14w,ref14x,ref14y}, Brans-Dicke theory \cite{ref14z}, Gauss-Bonnet theory \cite{ref15} and $f(R,T)$ theory \cite{ref16}, which are obtained by changing the geometric part of Einstein field equation. It is shown that within mimetic gravity, a theory of modified gravity which has gained a lot of interest recently, the coincidence problem is alleviated \cite{ref16a}.  The feasible solution of the cosmic coincidence problem can also be found by taking the interaction between the dark matter and the dark energy \cite{ref14s}. Also, more generally regarding interacting dark energy models, these models can have important observational consequences, particularly for solving the well-known $H_{0}$ tension \cite{ref16b,ref16c}.\\

	The HDE has the number of  significant features of the quantum gravity and has the traits of holographic principle \cite{ref17,ref18}, which states that degrees of freedom (the actual entropy of a system that has to be stored on some boundary in a holographic scenario)  are dependent on bounding area instead of volume. The reason for the flat FRW Universe was not known when HDE was considered in terms of Benkenstein entropy using infrared cut-off with the Hubble horizon \cite{ref19,ref20,ref21}. Physicists have taken various entropies with different cut-off scales like the interaction between cold dark matter and dark energy or combination of the mentioned approaches\cite{ref22,ref23}.\\
	
	In the literature \cite{ref14n,ref25,ref26,ref27}, HDE model has been considered widely and examined as $\rho _D \propto \Lambda ^4$, while relation between the IR cut-off L, UV cut-off $\Lambda$ and the entropy S is $\Lambda{^3} L{^3} \leq (S)^{\frac{3}{4}}$. So, the combination of the IR cut-off with the entropy gives the energy density of the HDE model.  
	The standard HDE model depends on Bekenstein-Hawking entropy $S = \frac{A}{4G}$, where $A = 4 \pi L^{2}$, thus the density is $\rho_{D}= \frac{3c^{2}}{8 \pi G}L^{-2}$,  where $c$ is numerical constant.   It must be stressed that this expression of  $\rho _D$ is obtained by combining the holographic principle and the
	dimensional analysis, instead of adding a DE term into the Lagrangian. Because of this extraordinary characteristic, HDE amazingly contrasts from some other theory of dark energy. The vacuum energy is associated with the UV cut-off and Ricci scalar, particle horizon, Hubble horizon,  event horizon,  etc. i.e. large scale structure of the Universe, is associated with the infrared (IR) cut-off. The HDE model endures the decision of IR cut-off problem. Numerous investigations of different (IR) cut-off's have been done in Refs. \cite{ref21,ref29,ref30,ref31,ref32,ref33}.\\     
	
	Various entropies are used for the investigation of the cosmological and gravitational incidence. The Tsallis HDE  \cite{ref35}, R$\acute{e}$nyi HDE \cite{ref36} and Sharma-Mittal HDE  \cite{ref37} are in demand and are extensively studied in literature. Differing from the usual HDE model with Bekenstein entropy, such models give a late-time accelerated Universe.  R$\acute{e}$nyi HDE depicts better stability as its own, in a non-interacting Universe \cite{ref36}. It is stable and Tsallis HDE \cite{ref38} is never stable if the Sharma-Mittal HDE becomes dominant in the Universe.
	So the inferences shows that R$\acute{e}$nyi  and Tsallis entropies can be obtained by Sharma-Mittal entropy \cite{ref40,ref41,ref42}. By considering the Hubble horizon as the IR cut-off, Tsallis HDE in Brans-Dicke cosmology has been studied \cite{ref43}, which demonstrates that both non-interacting and interacting cases are classically unstable.
	Recently Tsallis agegraphic dark energy model along with pressure-less dust was examined by Zadeh et al. \cite{ref44} and they observed that these models are classically unstable and show a late-time acceleration in non-interacting cases. 
	Investigation of Sharma-Mittal,  R$\acute{e}$nyi and Tsallis HDE, models have been done in \cite{ref45} by taking Loop Quantum Cosmology in consideration.\\

	 It is important to mention that observations admit
		an interaction between the dark sectors (DM and DE)
		of cosmos which can solve the coincidence problem
		and the tension in current observational values of the
		Hubble parameter \cite{ref45a,ref45b,ref45c,ref45d,ref45e}.
	
	HDE models generate late-time acceleration using infrared cut-off with the Hubble horizon when there is some interaction between dark energy and dark matter \cite{ref14n,ref47,ref48,ref49,ref50,ref50a,ref50b}. It can give late-time acceleration with matter-dominated decelerated expansion in the past.  Wei and Cai \cite{ref50c}, explored the interacting agegraphic dark energy (IADE) model considering the three most popular forms ( $Q=3\gamma H\rho_{tot}$, $3\beta H\rho_{m}$, $3\alpha H\rho_{q}$ \cite{ref45a,ref50c1,ref50c2}, where $\gamma$, $\beta$, $\alpha$ are constants) of interaction. The cosmological consequences of the recently proposed Tsallis HDE has been investigated by taking the interaction between DE and dark matter as, $Q=H(\alpha\rho_{m}+\beta\rho_{D})$ with IR cutoff as Hubble horizon  \cite{ref50d}.
		Also, considering various IR cut-off, the authors \cite{ref50e}, explored
		the THDE model evolution  and investigated
		their cosmological consequences with an interaction ( $Q=3b^{2}H(\rho_{m}+\rho_{D})$, where $b^{2}$ is coupling constant) between the dark matter and DE of
		the Universe with Hubble horizon as IR cut-off. Recently, The interaction rate between dark matter and DE has been
		reconstructed  for the HDE models by considering Hubble horizon as IR cut-off with three distinct forms of the interaction term $Q$ \cite{ref51}.\\  

		The scenario of interaction
		between DM and DE is one of such alternative models,
		which is the  subject interest of the present work. In this work, we investigate the evolution of
		our Universe by considering an interaction between the RHDE and DM
		whose IR cut-off is the Hubble horizon. The
		behaviour of the RHDE deceleration parameter,  the equation
		of state parameter EoS, the density parameter has been studied for the present
		model. Furthermore, we also investigate the stability
		of the RHDE model in
		the present scenario. However, the present work has some similarities and differences with other models reported by
		\cite{ref50c,ref50d,ref50e,ref51} in different ways. Firstly, in this paper, the
		functional form of the interaction term is different  from \cite{ref50c,ref50d,ref50e}, while similar
		to the interaction proposed in \cite{ref51}. Secondly, this work comprises of  the recently proposed  RHDE model from three different parametrizations of the interaction term $Q$, while in \cite{ref51}, the  HDE model is considered.\\

	The interaction function Q is supposed to be proportional to H$\rho_{D}$, where H is the Hubble parameter and $\rho_{D}$ is the R$\acute{e}$nyi HDE density. The strength of the interaction depends on the proportionality parameter $ \alpha$. Praseetha and Mathew checked at the apparent and event horizon in interacting holographic models whether the second law of thermodynamics is valid \cite{ref52}.\\
	
	These works are behind our motivation for investigating the cosmological consequence of R$\acute{e}$nyi HDE model by using infrared cut-off with the Hubble horizon and also by taking three different parametrizations of the interaction function Q, in the context of interacting flat FRW Universe. The organization of the paper is as follows: In Sect. 2, we discuss field equations in flat FRW Universe, the RHDE Model and calculated some cosmological parameters in the interacting RHDE model. In Sect. 3, we analysed the cosmological behaviour of the interacting RHDE for the model I, model II and model II. Finally, in the last section, we concluded outcomes.\\
	
	\section{Parametrization on the interaction term and RHDE}
	
	The metric for an isotropic and homogeneous flat FRW Universe is given by  :
	\begin{eqnarray}
	\label{eq1}
	ds^{2} = -dt^{2}+a^{2}(t)\Big(dr^{2} +  r^{2}d\Omega^{2}\Big),
	\end{eqnarray} 
	where a(t) is known as the scale factor. The Hubble parameter is determined as, $ H = \frac{\dot a}{a} $, where the dot represents derivative with respect to cosmic time.
	The Friedmann equations, in the form of Hubble parameter are given as, 
	
	\begin{eqnarray}
	\label{eq2}
	H^2 = \frac{1}{3} (8 \pi  G) \left(\rho_D+\rho _M\right),
	\end{eqnarray} 
	where $\Omega_{D} = \frac{1}{3}8 \pi  G\rho_{D}H^{-2} $ and  $\Omega_{m} = \frac{1}{3}8 \pi  G\rho_{m}H^{-2} $  are the energy density parameters of the RHDE and pressure less matter,  respectively, expressed as fractions of critical density $\rho_{c}= \frac{3H^{2}}{8 \pi  G}$. Also, $\rho_{m}$ and $\rho_{D}$ denote the energy density of matter and RHDE, respectively, and $\frac{\rho_{m}}{\rho_{D}} = r$ represents the energy density ratio of two dark components \cite{ref22,ref54}.
	Now Eq. (\ref{eq2}) can be written as:
	\begin{eqnarray}
	\label{eq3}
	1  = \Omega _D+\Omega _m, 
	\end{eqnarray}
	The conservation law to the interacting RHDE and matter are found as :
	\begin{eqnarray}
	\label{eq4}
	\dot \rho_{m} + 3 H \rho_{m} = Q,
	\end{eqnarray}
	\begin{eqnarray}
	\label{eq5}
	\dot \rho_{D} + 3H (\rho_{D} + p_{D}) = - Q,
	\end{eqnarray}
	Here Q denotes the interaction function and  $ \omega _D = p _D/\rho _D$ gives the equation of state. Equations \ref{eq4} and \ref{eq5} become decoupled for $Q = 0$ permitting the autonomous conservation of dark matter and dark energy.  The mutual interaction implies a decaying of the holographic energy component into CDM \cite{ref54a,ref54b,ref54c,ref54d,ref54e}.  Kim et al. \cite{ref54a}, proposed that if there exists a source/sink in the right-hand side of the continuity equation, one must be careful to define the EoS. In this case the effective EoS is the only candidate to represent the state of the mixture of two components arising from decaying of the holographic energy into CDM. This is quite different from the non-interacting case. However, it was suggested that  one  have  to  use $\omega_{D}^{eff}$ when considering the interaction \cite{ref54a}. In this study, we have taken three different parametrizations of the interaction function Q. The common form of the interaction function is taken to be  $ Q = 3 \ \alpha (z)\  H \ \rho_{D}$, where $ \alpha$ represents the coupling term which is a function of redshift $z$. The rate of interaction between dark matter and dark energy is obtained as $\Gamma=\frac{Q}{\rho_{H}}$ \cite{ref54a,ref54f} and thus it can be expressed as
	
	\begin{eqnarray}
	\Gamma=3H(z)\alpha(z).
	\end{eqnarray}
	The rate of energy transfer and also the direction of energy flow depend on this term. The dark energy equation of state parameter $\omega_{D}$ is related to the effective equation of state parameter as \cite{ref54a}
	
	\begin{eqnarray}
	\omega_{eff}= \Omega_{D}\omega_{D}
	\end{eqnarray}
	
	Now the coincidence parameter $(r)$ is defined as $r = \rho _m/\rho _D$, which is constant in case of HDE in a spatially flat Universe with Hubble horizon as the IR cut-off \cite{ref55}.
	The behaviour of this ratio is crucial for the ``conventional'' form of the ``coincidence
	problem'', namely: ``why are the  dark energy and matter  densities of precisely
	the same order today”? In principle, dark energy and matter redshift at different rates \cite{ref55a1}. In \cite{ref55a1}, it is concluded  that there exists a preferred class of DE models for which the dynamics of the energy density ratio is entirely determined by the spatial curvature. For vanishing curvature,
	the energy density ratio remains constant. This feature is highlighted in a broader context
	and demonstrate that a constant or slowly varying (as the consequence of a non-vanishing spatial curvature) energy density ratio is compatible with a transition from decelerated to
	accelerated expansion under the condition of a growing interaction parameter. These models are singled out by a dependence
	$\rho_{D}\propto H^{2}$ where $\rho_{D}$ is the dark energy density.
	Exactly this dependence is characteristic for a certain type of dark energy models, inspired
	by the holographic principle \cite{ref18}. Also, a different way to understand the dynamics of the ratio  $r$ can be seen in details \cite{ref55a1}.\\

	In this study to reconstruct the interaction function Q, we have taken three different ansatzes which are given in \cite{ref51} as \\
	
	Model (I) 
	\begin{eqnarray}
	\label{eq6}
	\alpha (z) =  \alpha_1 +  \alpha_2 (1 + z),
	\end{eqnarray}
	
	Model (II) 
	\begin{eqnarray}
	\label{eq7}
	\alpha (z) =  \alpha_1 +  \alpha_2 \left(\dfrac{ z}{1 + z}\right),
	\end{eqnarray}
	
	Model (III) 
	\begin{eqnarray}
	\label{eq8}
	\alpha (z) =  \alpha_1 +  \alpha_2 \left(\dfrac{ 1}{1 + z}\right),
	\end{eqnarray}										
	
	Where $ \alpha_1$ and $ \alpha_2$ are constant parameters. The model I, II and III have a linear, mixed and inverse dependence on $z$. So, the aforementioned three models lead us to a non-interacting case after reduction, if $ \alpha_1$ and $ \alpha_2$ are taken as zero. Here we find two parameters  $\alpha _1$  and  $\alpha _2$ since for these three models $r$ is varying. In \cite{ref51}, the distance modulus measurements of type Ia supernova from the Joint Light-curve Analysis (JLA) \cite{ref57} and the observational measurements of Hubble parameter (OHD) have been used to constrain these model
	parameters for holographic dark energy models. Cosmic Chronometer method  \cite{ref58}, measurements from galaxy distribution \cite{ref59} and from Lymann $- \alpha$ forest distribution \cite{ref60} methods are used to measure the OHD.
	 In this work, we use different  values of $\alpha _1$  and  $\alpha _2$ to see the models behaviour since the interaction function $Q$ depends on the parameters $ \alpha_1$ and $ \alpha_2$. A negative $Q$ shows the flow of energy from the dark matter to the dark energy and a positive $Q$ shows the reverse.\\
	\begin{figure}
		\begin{center}
			(a)\includegraphics[width=5cm, height=5cm ]{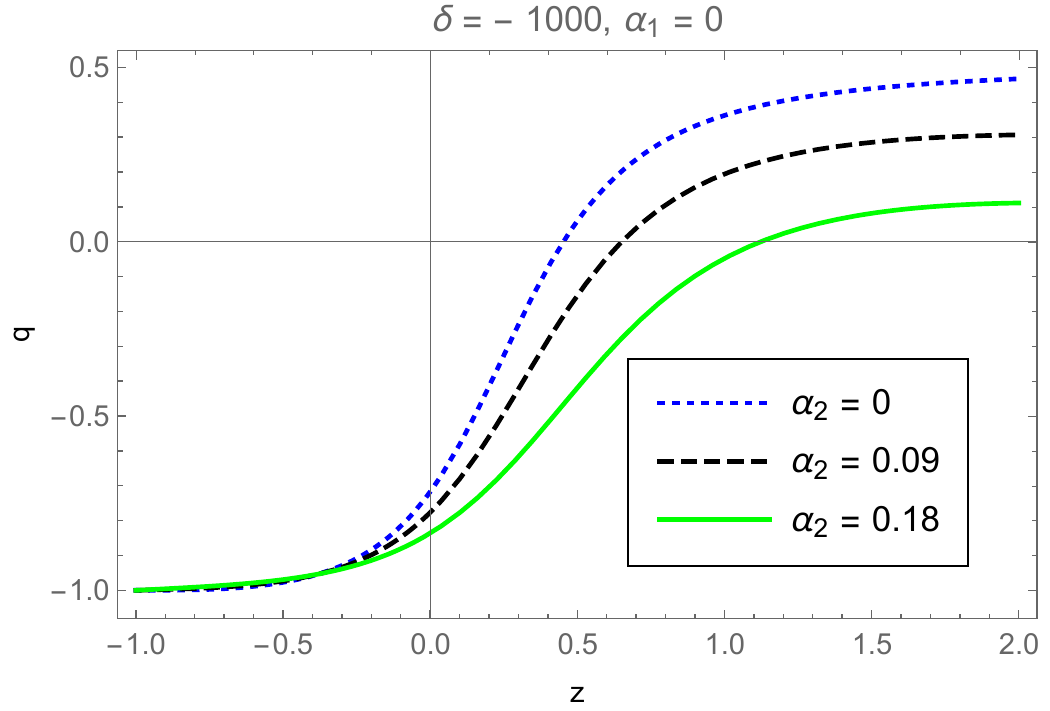}
			(b)\includegraphics[width=5cm, height=5cm]{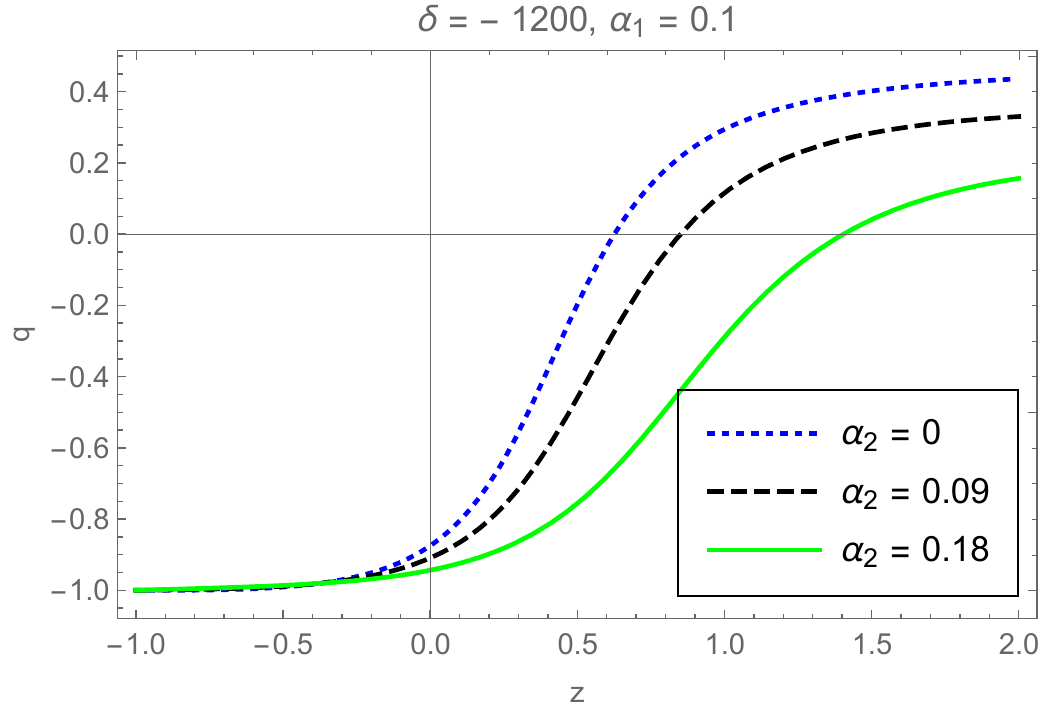}
			(c)\includegraphics[width=5cm, height=5cm]{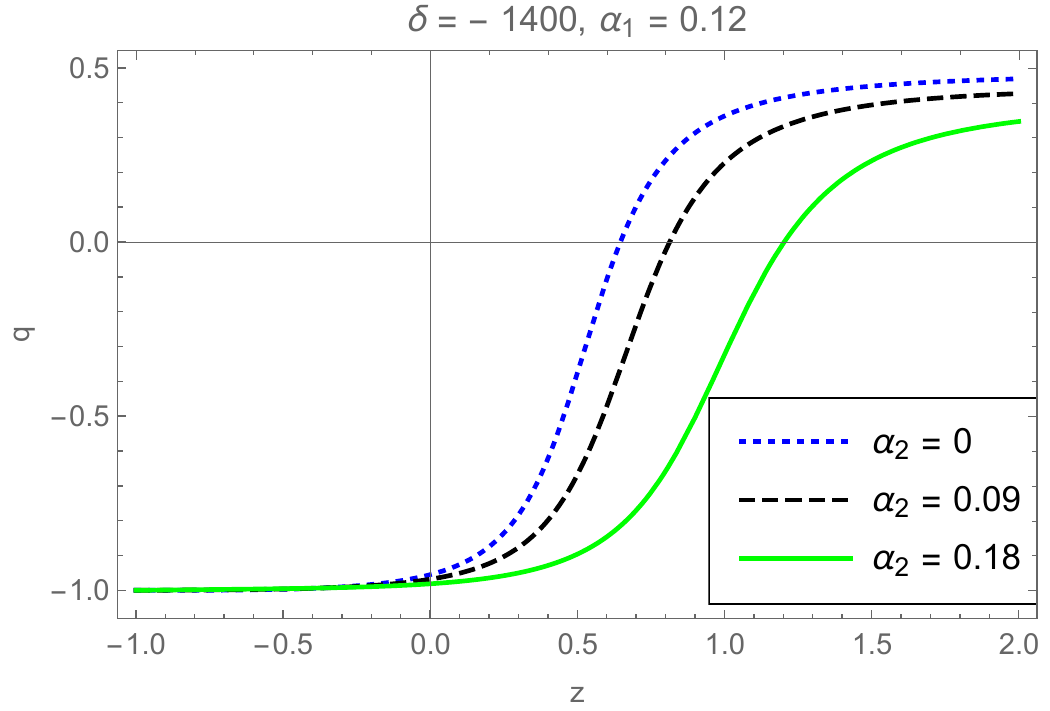}
			
			\caption { The evolution of deceleration parameter ($q$) in the RHDE model (I) versus redshift $z$ for different values of model parameter $\delta$, $\alpha _1$, and $\alpha _2$ where $H_0$ = 69.6, $\Omega _{D0}$ = 0.70.}
			
		\end{center}
	\end{figure}
	
	The form of the Bekenstein entropy of a system is $S = \frac{A}{4}$ , where $A = 4 \pi L^{2}$ and L is the IR cut-off. Another modified form of the R$\acute{e}$nyi entropy \cite{ref36} is given as: 
	
	\begin{eqnarray}
	\label{eq9}
	S=\frac{1}{\delta}\log  \left(\frac{\delta}{4} A + 1\right ) = S=\frac{1}{\delta}\log  \left(\pi\delta L^{2}+ 1\right ),
	\end{eqnarray} 
	R$\acute{e}$nyi HDE density, by considering the assumption $ \rho _d\  dV \  \propto \   T dS$, takes the following form:
	
	\begin{eqnarray}
	\label{eq10}
	\rho _D=\frac{3 c^2}{8 \pi L^{2}} (\pi\delta L^{2}+ 1)^{-1},
	\end{eqnarray} 
	By taking Hubble horizon as an IR cut-off $L = \dfrac{1}{H}$, we obtained:
	\begin{eqnarray}
	\label{eq11}
	\rho _D=\frac{3 c^2 H^2}{8 \pi  \left(\frac{\pi  \delta }{H^2}+1\right)},
	\end{eqnarray} 
	where $ c^{2} $ is a numerical constant as usual. Now, by the definition of $r$, we obtain \cite{ref54a}
	
	\begin{eqnarray}
	\label{eq11a}
	r= \frac{\rho_{m}}{\rho_{D}} = \frac{\frac{1}{3} \rho_{m}M_{p}^{-2}H^{-2}}{ \frac{1}{3}\rho_{D}M_{p}^{-2}H^{-2}}= \frac{\Omega_{m}}{\Omega_{D}} = \frac{1-\Omega_{D}}{\Omega_{D}}
	\end{eqnarray}
	\begin{eqnarray}
	\label{eq12}
	r=\frac{1}{\Omega _D}-1.
	\end{eqnarray}
	For the flat Universe, we have $1-\Omega_{D}=\Omega_{m}$. Moreover, in this case $r$ stays varying.  Now, taking the time derivative of Eq. (\ref{eq11}), we get
	\begin{eqnarray}
	\label{eq14}
	\dot{\rho _D} =2  \frac{\dot{H}}{H} \rho _D \left(\frac{\pi  \delta }{\pi  \delta +H^2}+1\right),
	\end{eqnarray}

	by taking the time derivative of Eq. (\ref{eq2}) and using the Eq. (\ref{eq4}) and Eq. (\ref{eq14}), and combining the result with Eq. (\ref{eq3}), we obtain
	\begin{eqnarray}
	\label{eq12}
	\frac{\dot{H}}{H^2}  =  -\frac{3}{2} (1 + \Omega _D \omega _D),
	\end{eqnarray}
	Now using Eq. (\ref{eq12}) we get  deceleration parameter $q$
	\begin{eqnarray}
	\label{eq13}
	q=   -1-  \frac{\dot{H}}{H^2}= \frac{1}{2} \left(3 \omega _D \Omega _D+1\right),
	\end{eqnarray}
	
	Now substituting  Eq. (\ref{eq14}) in  Eq.(\ref{eq5}). We get EoS parameter as: 
	\begin{eqnarray}
	\label{eq15}
	\omega _D= -1 -\alpha -\frac{2 \pi  \delta  \dot{H}}{3 H^2 \left(\pi  \delta +H^2\right)}-\frac{2 \dot{H}}{3 H^2},
	\end{eqnarray}

	Taking time derivative of  $\Omega_{D} = \frac{8 \pi  G  \rho_D}{3 H^{2}}$  and by using Eq.(\ref{eq14}), we get
	
	\begin{eqnarray}
	\label{eq16}
	\dot{\Omega _D}=\frac{2 \pi  \delta  \dot{H} \Omega _D}{H \left(\pi  \delta +H^2\right)},
	\end{eqnarray}
	
	Finally, we explore the stability of the RHDE model as:
	
	\begin{eqnarray}
	\label{eq17}
	v_s^2=\frac{{dp}_D}{{d\rho }_D}=\frac{\rho _D}{{\dot\rho _D}}  {\dot\omega _D} + \omega _D
	\end{eqnarray}
	
	Physical values of the dark energy squared sound speed should lie in the
	region $0\leq v^2_s\leq 1$. Outside of this region, one finds gradient /
	tachyonic instabilities and/or instabilities connected to
	superluminal propagation. In the standard scenario, one
	fixes  $v^2_s = 1$. Hence, it is required to check what is the impact
	of reducing the dark energy squared sound speed to values as low as
	$0$, and whether the effects on the CMB can be mimicked
	by dark energy baryon scattering \cite{ref55a}. One important point to note
	is that to study the effect of the DE squared sound speed,
	one must again consider values of the DE EoS $\omega_{D} \neq -1$ since
	a cosmological constant has no perturbations and hence a
	cosmology with DE in the form of a cosmological constant
	has no sensitivity to the DE sound speed. The impact of
	the DE sound speed on the CMB power spectrum has been
	discussed in detail in \cite{ref55b}. There it was
	found that the effect of decreasing $v^2_s$ from $1$ to $0$ is to suppress
	the late-time integrated Sachs-Wolfe (LISW) effect when the DE EoS satisfies $ \omega_{D} >-1$,
	but for $ \omega_{D} <-1$, decreasing the sound speed results in an
	enhancement of the LISW effect. Heuristically, at least for
	the case where $\omega_{D} > -1$, this can be understood as follows:
	the more we decrease $v^2_s$, the more DE can cluster and effectively
	behave as ``cold" dark energy. Clustering enhances the DE perturbations, which protect the
	potentials from decaying, thus leading to a smaller contribution
	to the late-time integrated Sachs-Wolfe (LISW) effect. We
	refer the reader to \cite{ref55b} for a mathematically
	rigorous discussion of the effect of $v^2_s$ on the CMB.
	See also e.g. \cite{ref55c,ref55d,ref55e,ref55f,ref55g,ref55h,ref55i} for further works examining
	the effect of $v^2_s$ on cosmological observations.

	\section{ Cosmological behaviour of the  interacting RHDE }

	In this section, we investigate the cosmological
	behaviour of the deceleration parameter $(q)$, equation of state parameter EoS ($\omega_{D}$), total energy density parameter ($\Omega_{D}$) and squared sound speed ($v_s^2$) via three ansatzes mentioned above for the parametrization of interaction term for different values of the R$\acute{e}$nyi parameter $\delta$. 
	In the following subsections we consider each model, in turn and  elaborate parameters numerically, imposing the initial conditions $H(z=0)$ = 69.6, $\Omega _{D}(z=0)$ = 0.70.
	
	\subsection{Model I}
	
	\begin{figure}
		\begin{center}
			(a)\includegraphics[width=5cm, height=5cm]{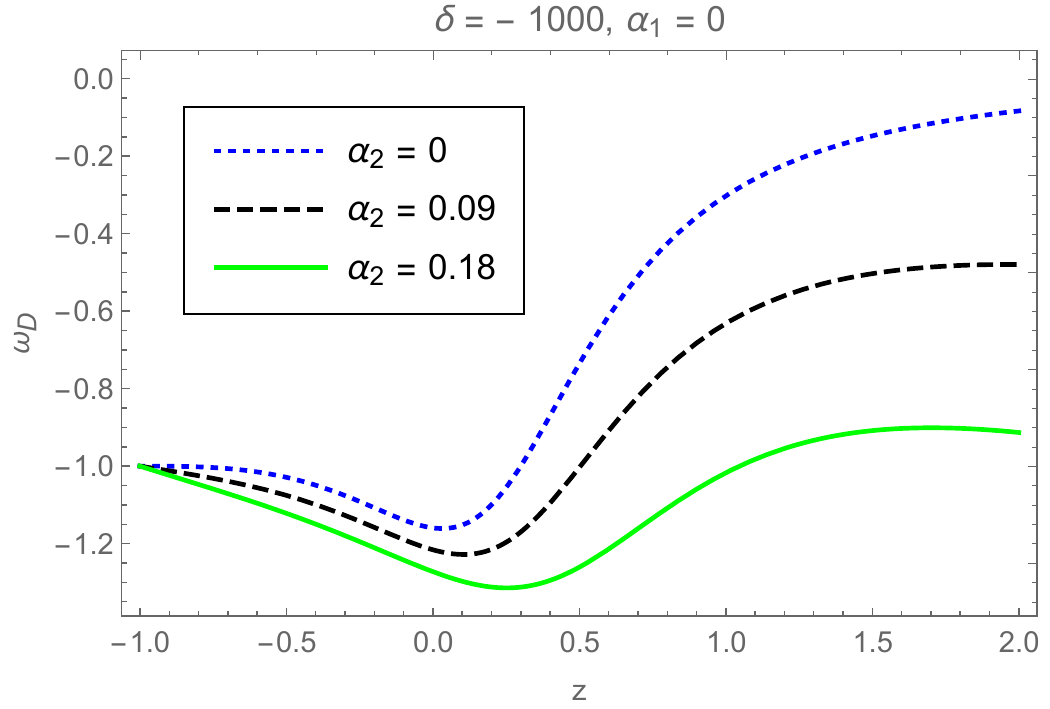}
			(b)\includegraphics[width=5cm, height=5cm]{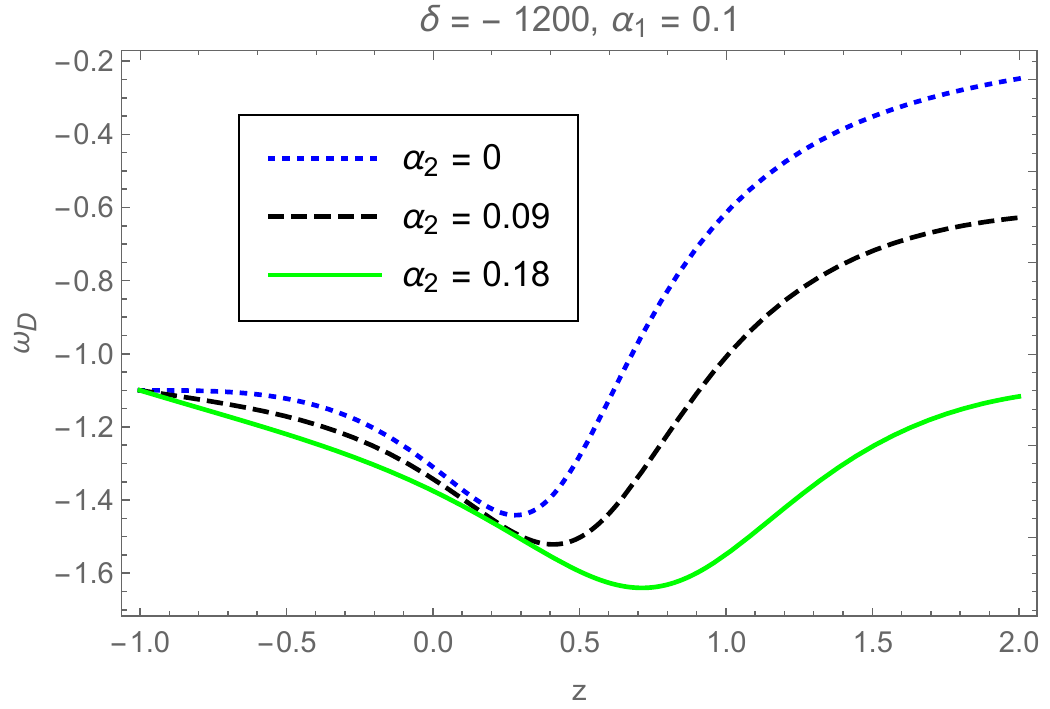}
			(c)\includegraphics[width=5cm, height=5cm]{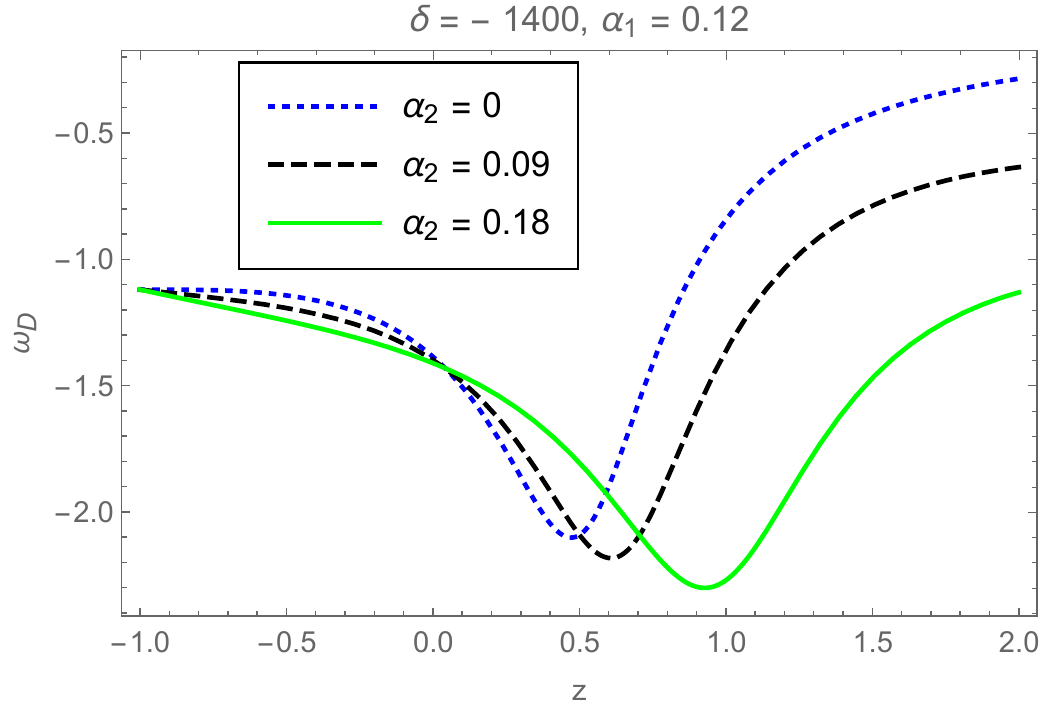}
			
			\caption { The evolution of EoS parameter $\omega _D$ in RHDE model (I)  versus redshift $z$ for different values of model parameter $\delta$, $\alpha _1$, and $\alpha _2$ where $H_0$ = 69.6, $\Omega _{D0}$ = 0.70.}
			
		\end{center}
	\end{figure}
	
	\begin{figure}
		\begin{center}
			(a)\includegraphics[width=5cm, height=5cm]{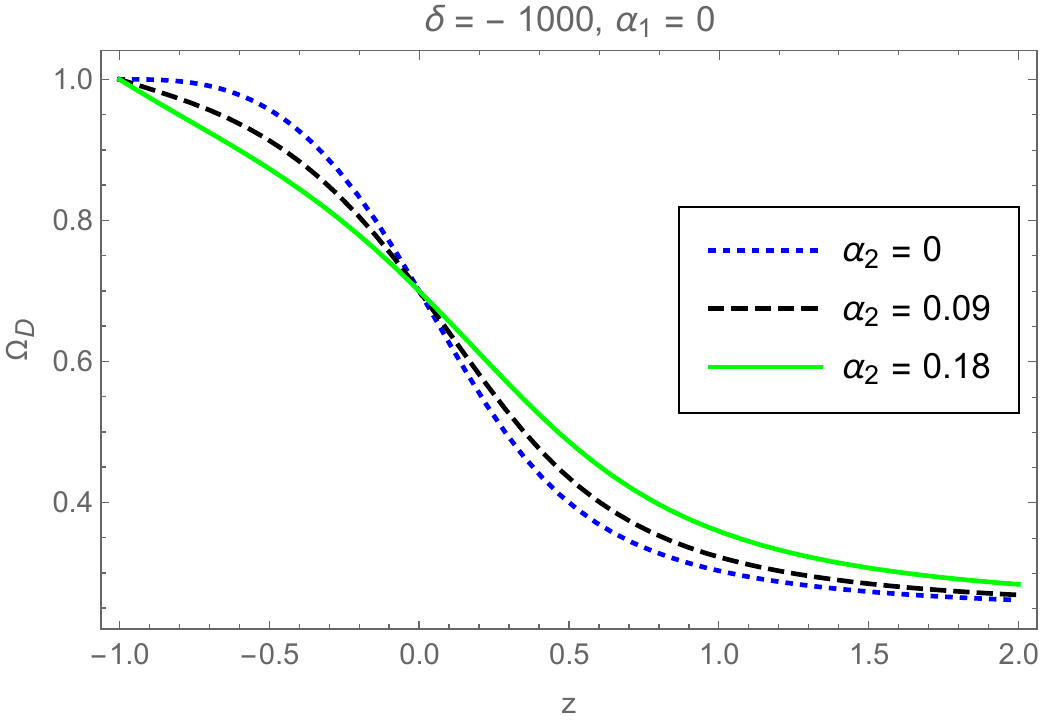}
			(b)\includegraphics[width=5cm, height=5cm]{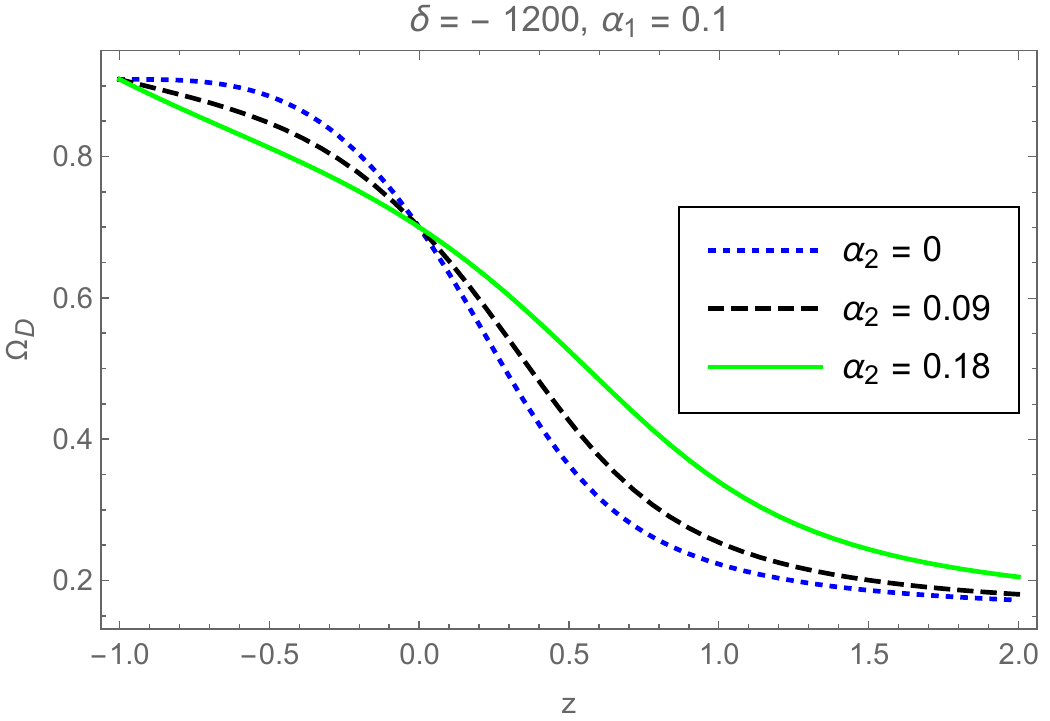}
			(c)\includegraphics[width=5cm, height=5cm]{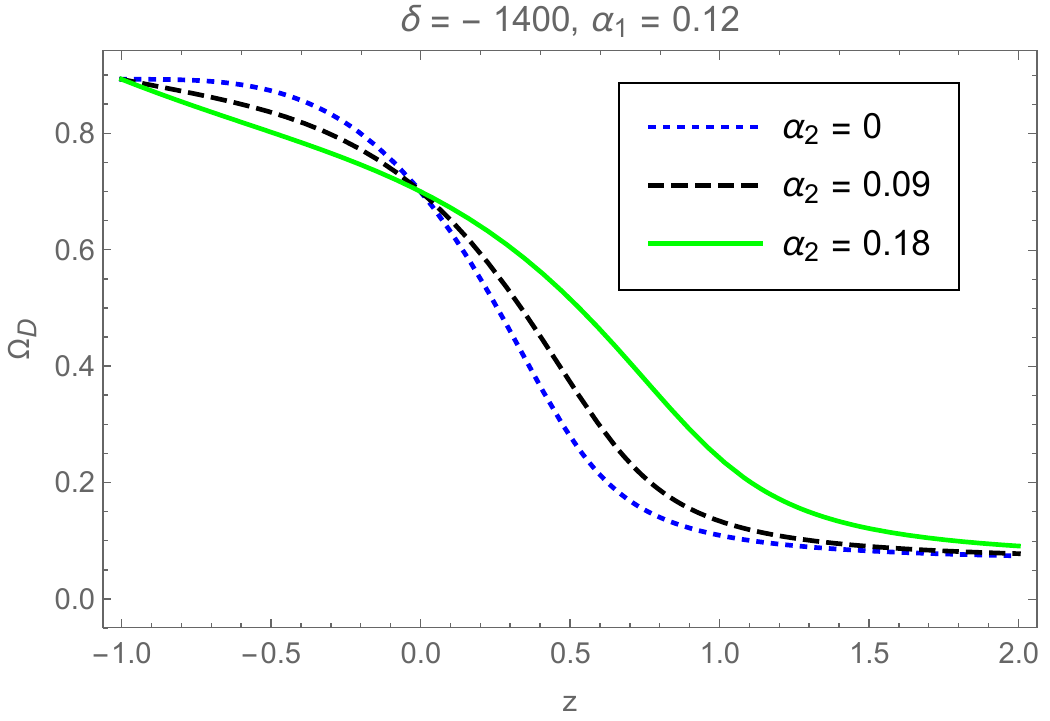}
			
			\caption { The evolution of energy density parameter  $\Omega _D$ in RHDE model (I) versus redshift $z$ for different values of model parameter $\delta$, $\alpha _1$, and $\alpha _2$ where $H_0$ = 69.6, $\Omega _{D0}$ = 0.70.}

		\end{center}
	\end{figure}
	
	\begin{figure}
		\begin{center}
			(a)\includegraphics[width=5cm, height=5cm]{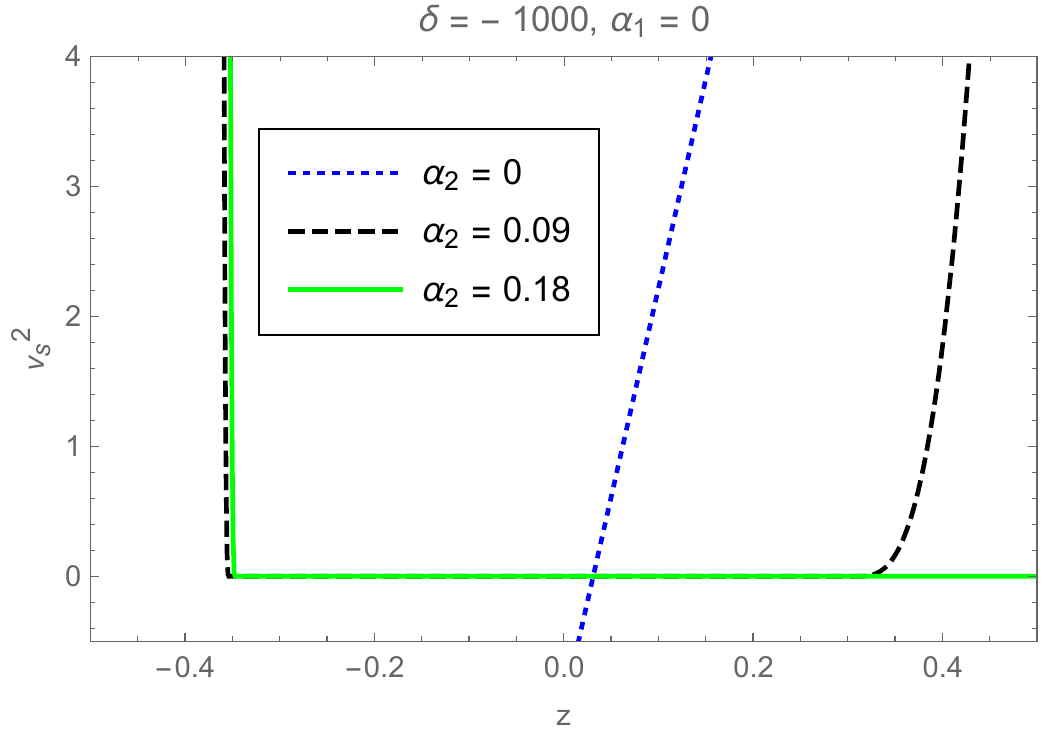}
			(b)\includegraphics[width=5cm, height=5cm]{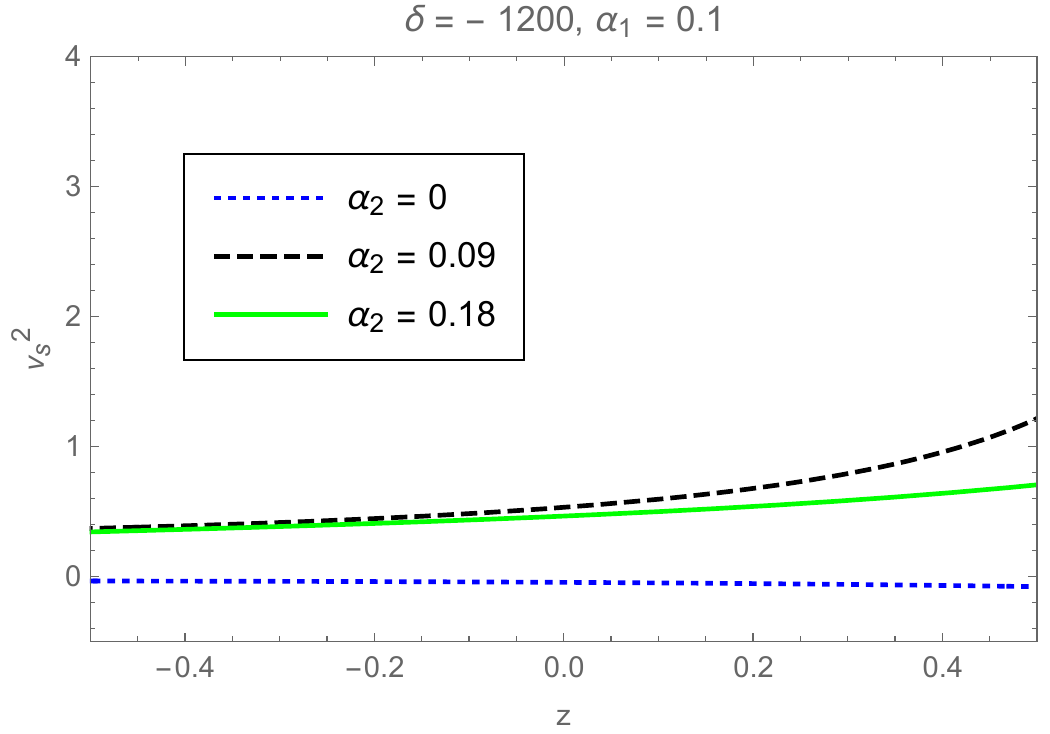}
			(c)\includegraphics[width=5cm, height=5cm]{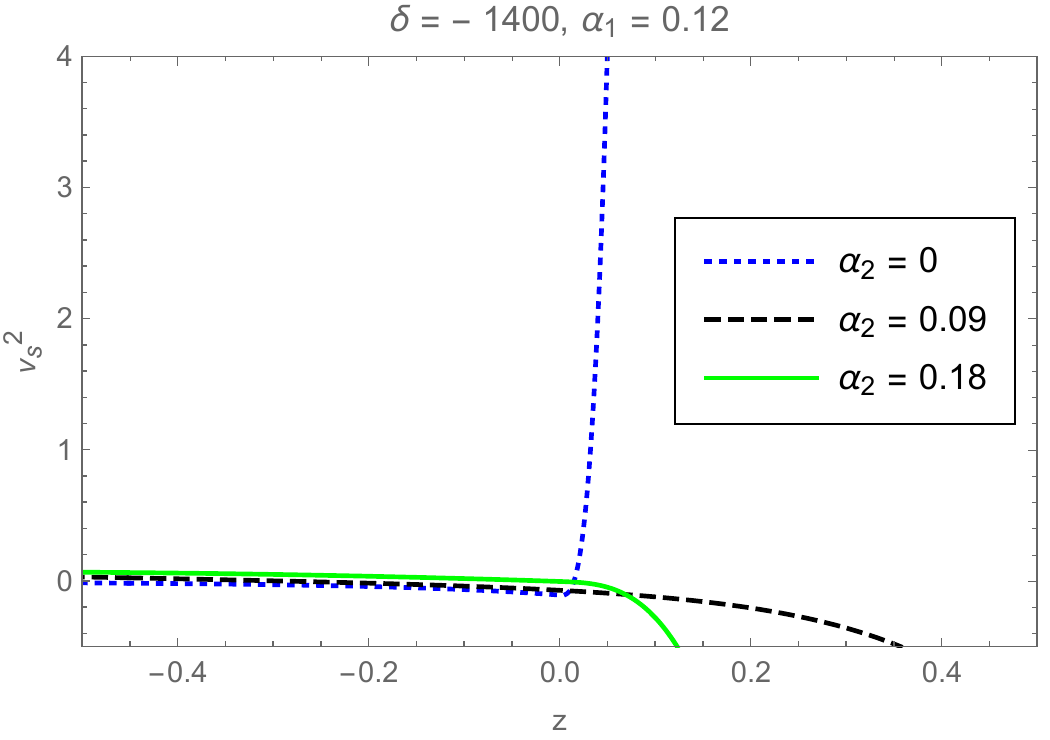}
			
			\caption { The evolution of square of the sound speed parameter $v_s^2$ in RHDE model (I)  versus redshift $z$ for different values of model parameter $\delta$, $\alpha _1$, and $\alpha _2$ where $H_0$ = 69.6, $\Omega _{D0}$ = 0.70.}

		\end{center}
	\end{figure}

	\begin{figure}
		\begin{center}
			(a)\includegraphics[width=5cm, height=5cm]{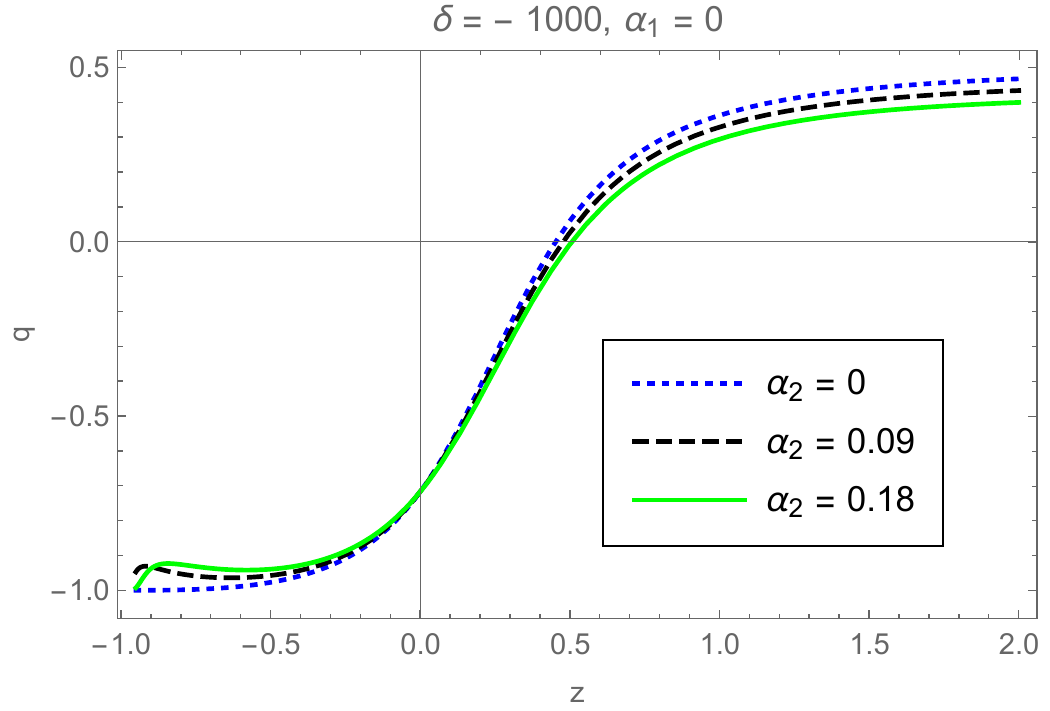}
			(b)\includegraphics[width=5cm, height=5cm]{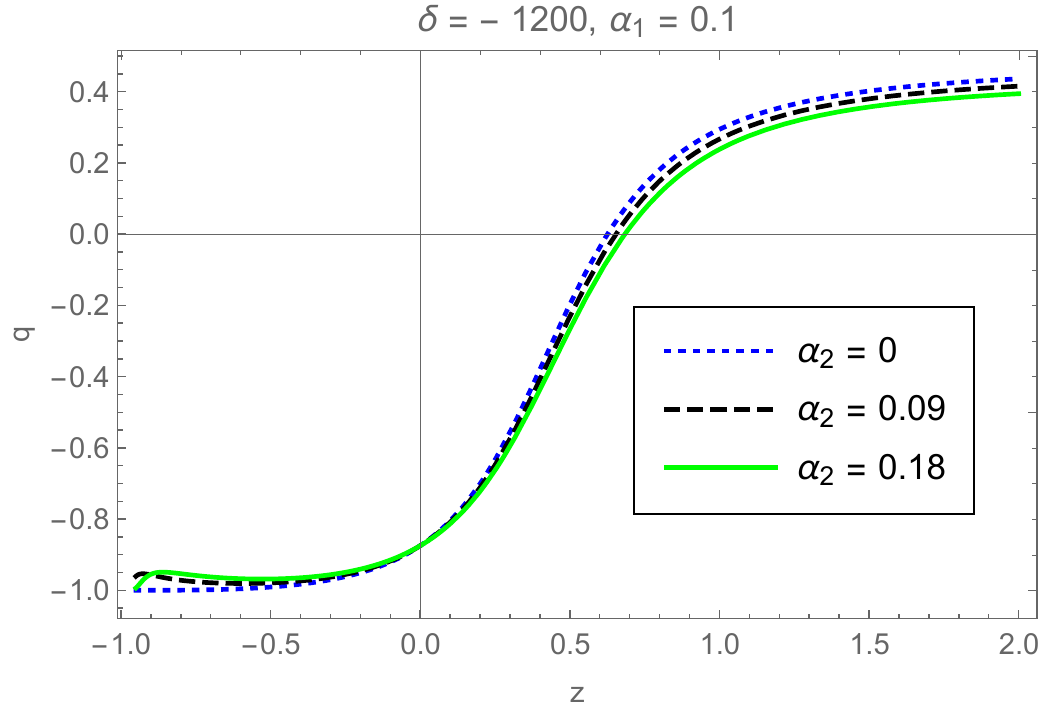}
			(c)\includegraphics[width=5cm, height=5cm]{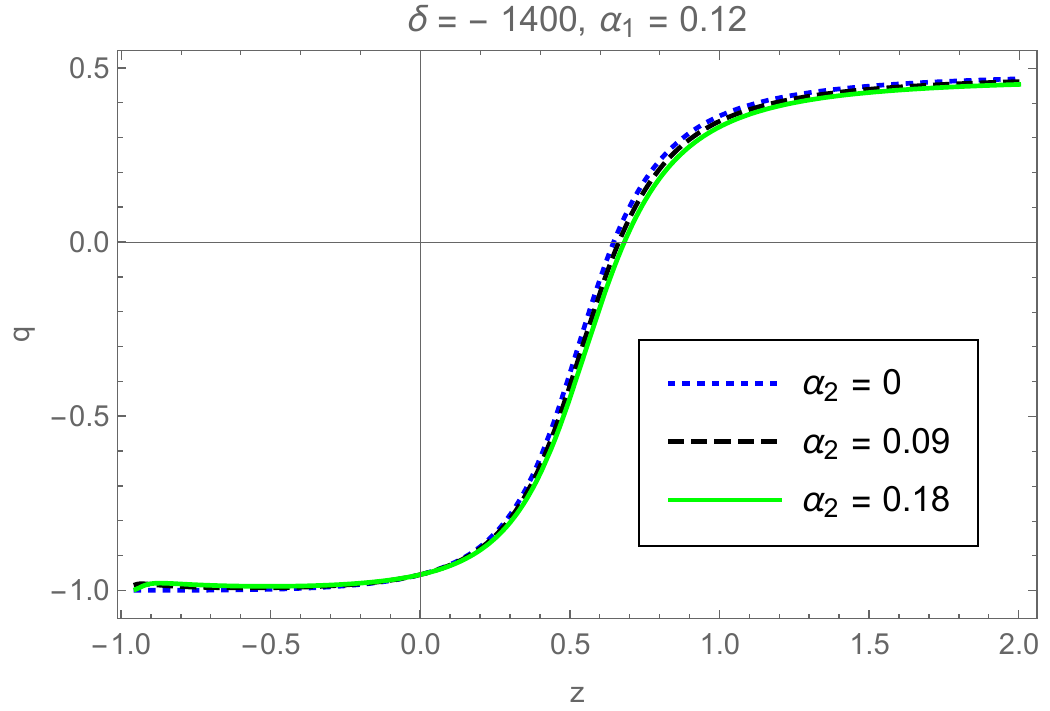}
			
			\caption { The evolution of deceleration parameter ($q$)  in the RHDE model (II)  versus redshift $z$ for different values of model parameter $\delta$, $\alpha _1$, and $\alpha _2$ where $H_0$ = 69.6, $\Omega _{D0}$ = 0.70.}
			
		\end{center}
	\end{figure}
	\begin{figure}
		\begin{center}
			(a)\includegraphics[width=5cm, height=5cm]{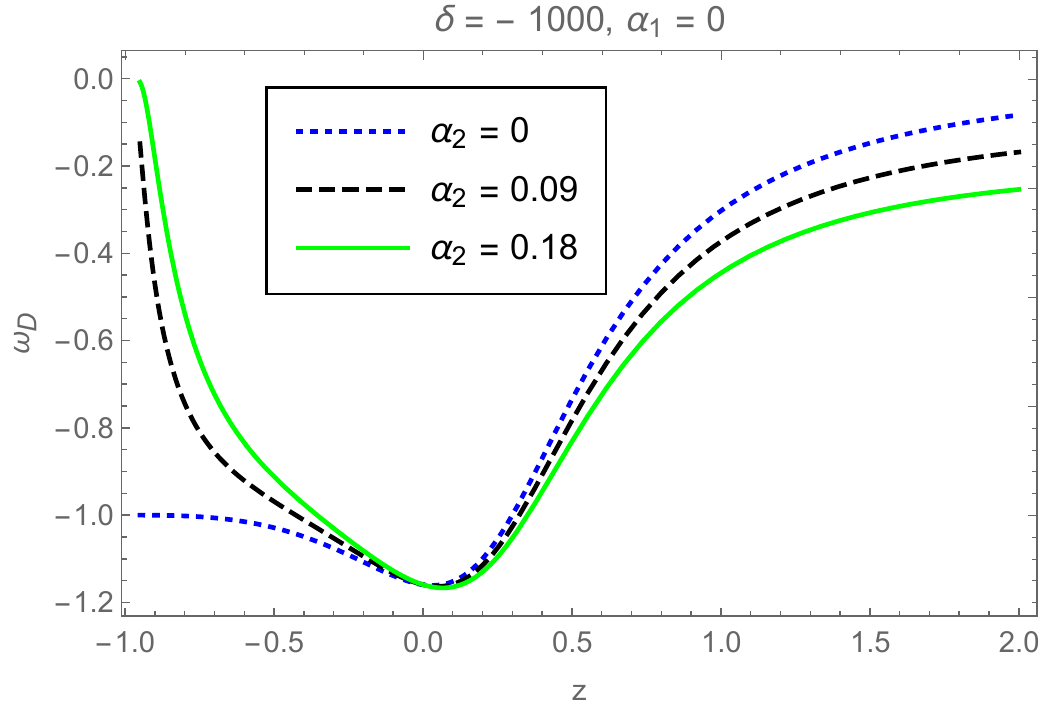}
			(b)\includegraphics[width=5cm, height=5cm]{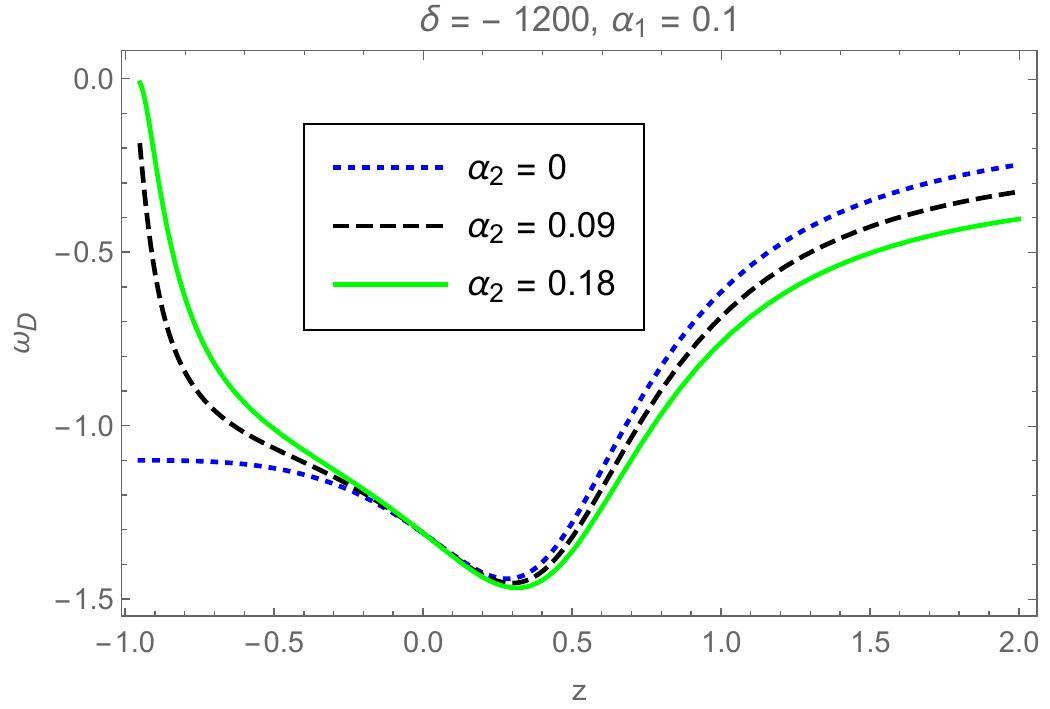}
			(c)\includegraphics[width=5cm, height=5cm]{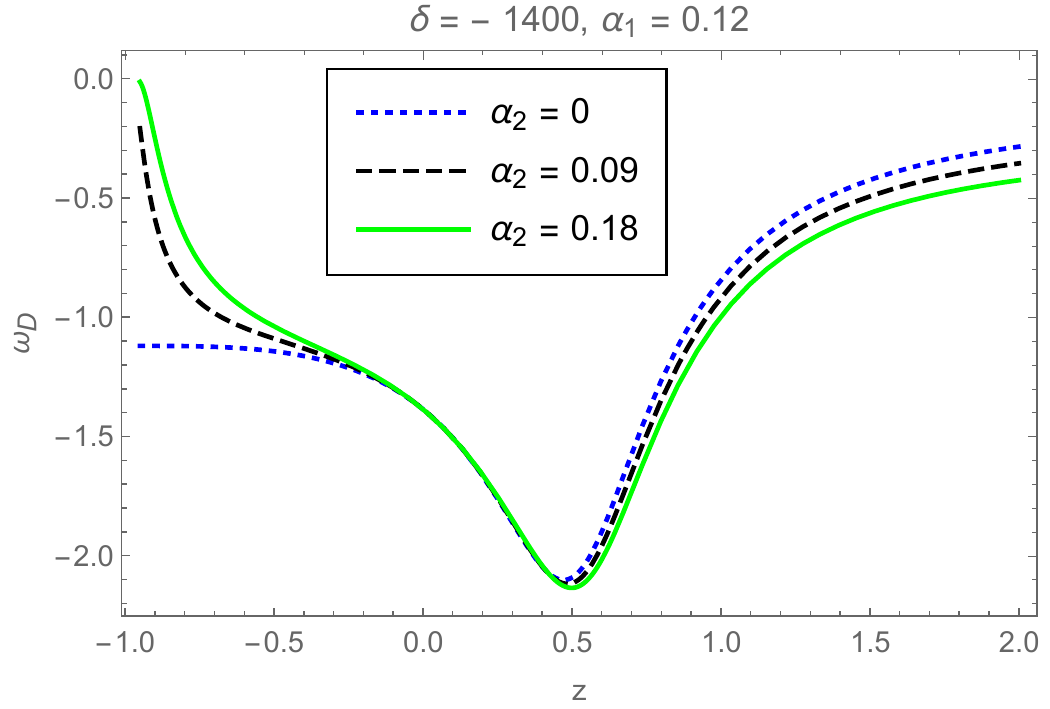}
			
			\caption { The evolution of EoS parameter $\omega _D$ in RHDE model (II) versus redshift $z$ for different values of model parameter $\delta$, $\alpha _1$, and $\alpha _2$ where $H_0$ = 69.6, $\Omega _{D0}$ = 0.70.}
			
		\end{center}
	\end{figure}
	
	\begin{figure}
		\begin{center}
			(a)\includegraphics[width=5cm, height=5cm]{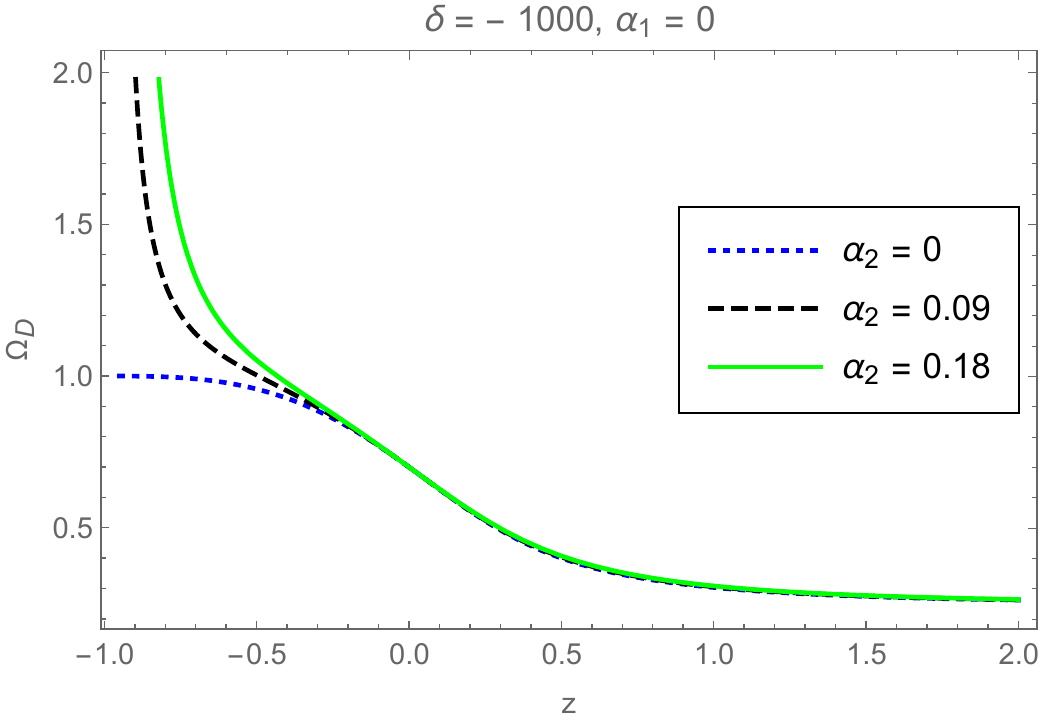}
			(b)\includegraphics[width=5cm, height=5cm]{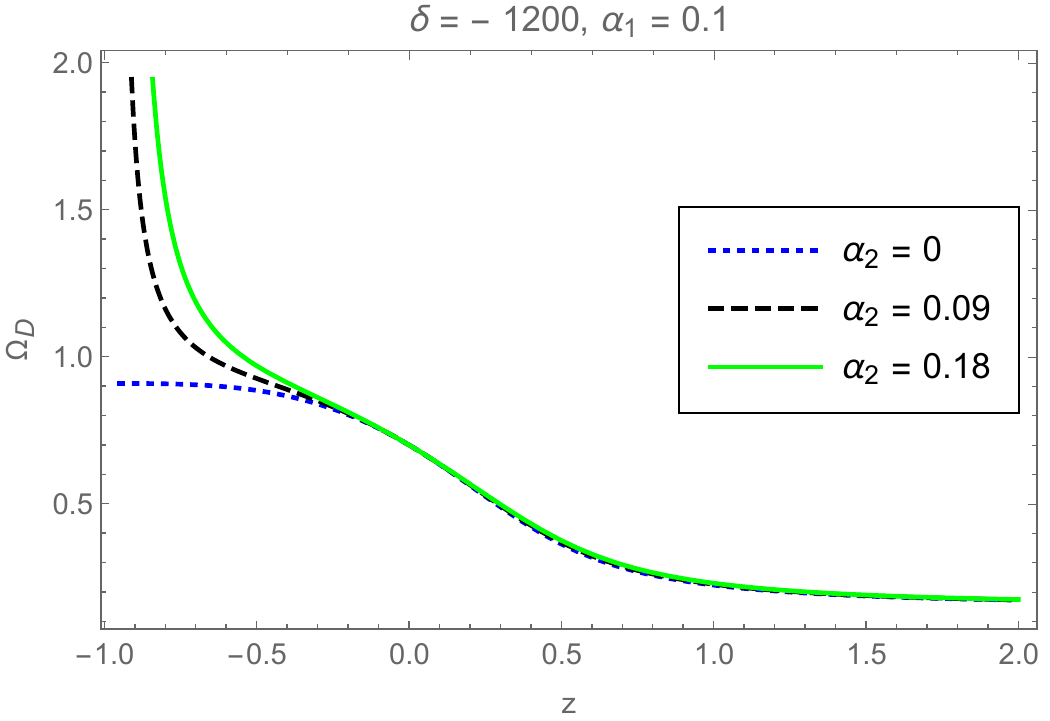}
			(c)\includegraphics[width=5cm, height=5cm]{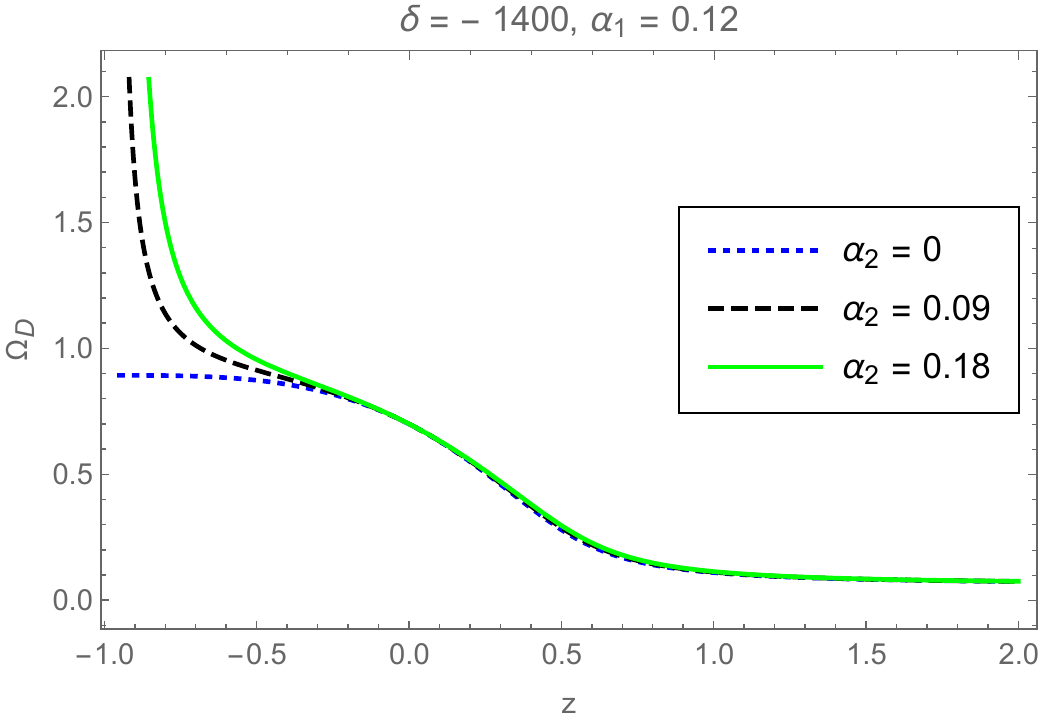}
			
			\caption { The evolution of energy density parameter  $\Omega _D$ in RHDE model (II) versus redshift $z$ for different values of model parameter $\delta$, $\alpha _1$, and $\alpha _2$ where $H_0$ = 69.6, $\Omega _{D0}$ = 0.70.}
			
		\end{center}
	\end{figure}
	
	\begin{figure}
		\begin{center}
			(a)\includegraphics[width=5cm, height=5cm]{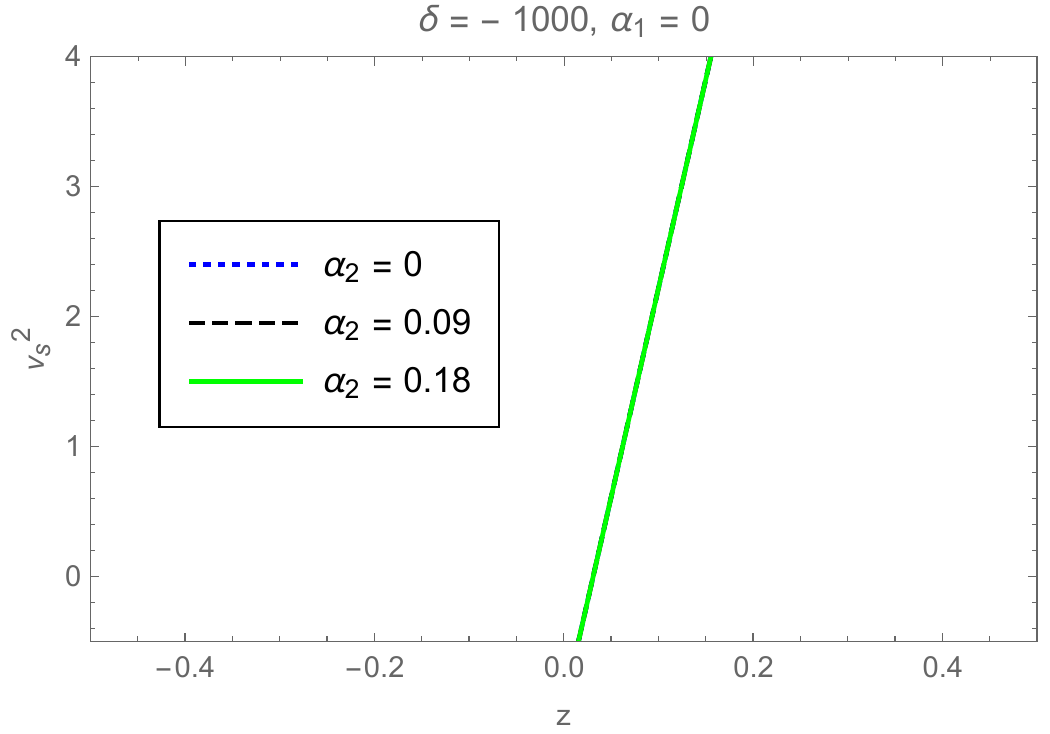}
			(b)\includegraphics[width=5cm, height=5cm]{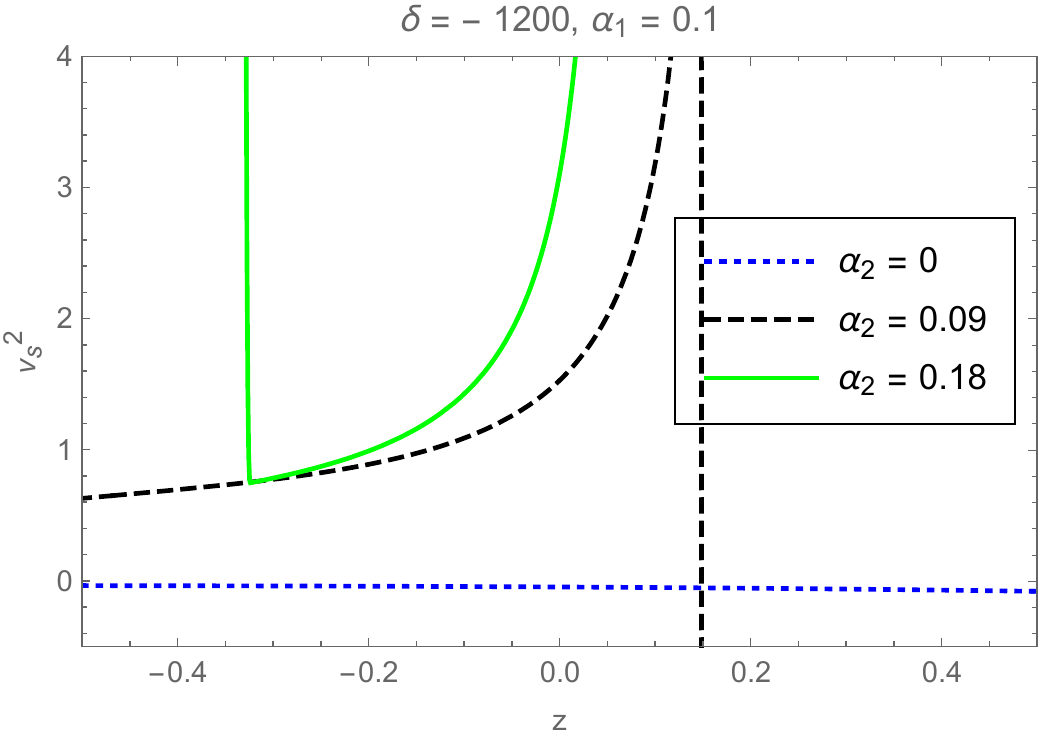}
			(c)\includegraphics[width=5cm, height=5cm]{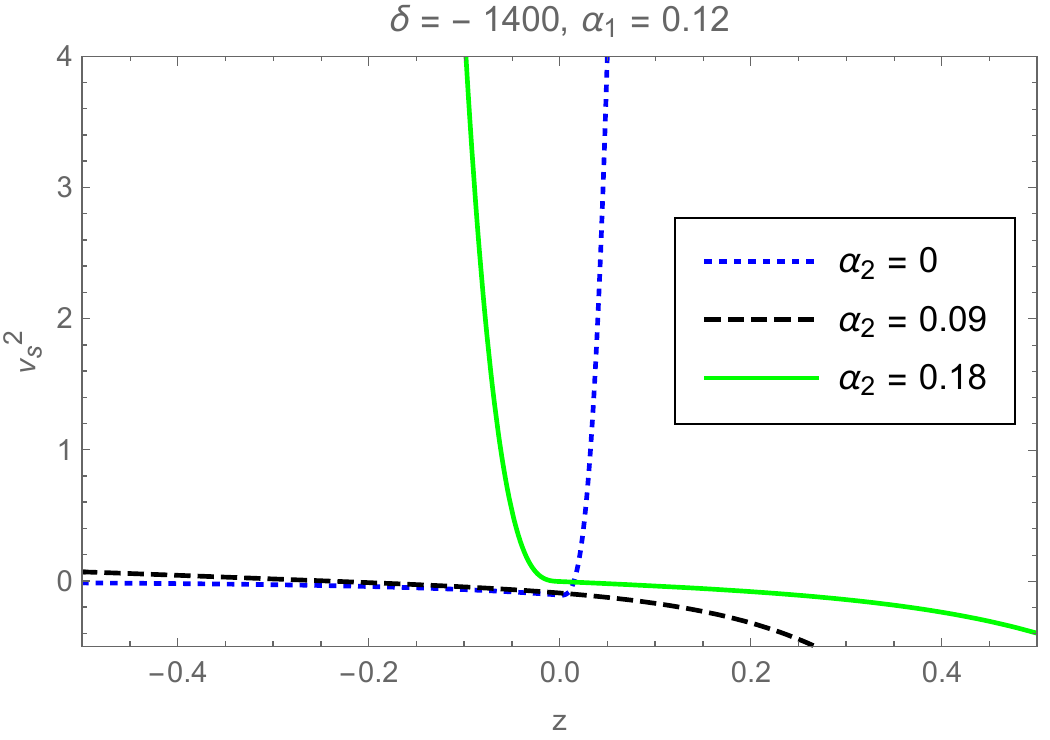}
			
			\caption { The evolution of square of the sound speed parameter $v_s^2$ in RHDE model (II)  versus redshift $z$ for different values of model parameter $\delta$, $\alpha _1$, and $\alpha _2$ where $H_0$ = 69.6, $\Omega _{D0}$ = 0.70.} 		
		\end{center}
	\end{figure}

	\begin{figure}
		\begin{center}
			(a)\includegraphics[width=5cm, height=5cm]{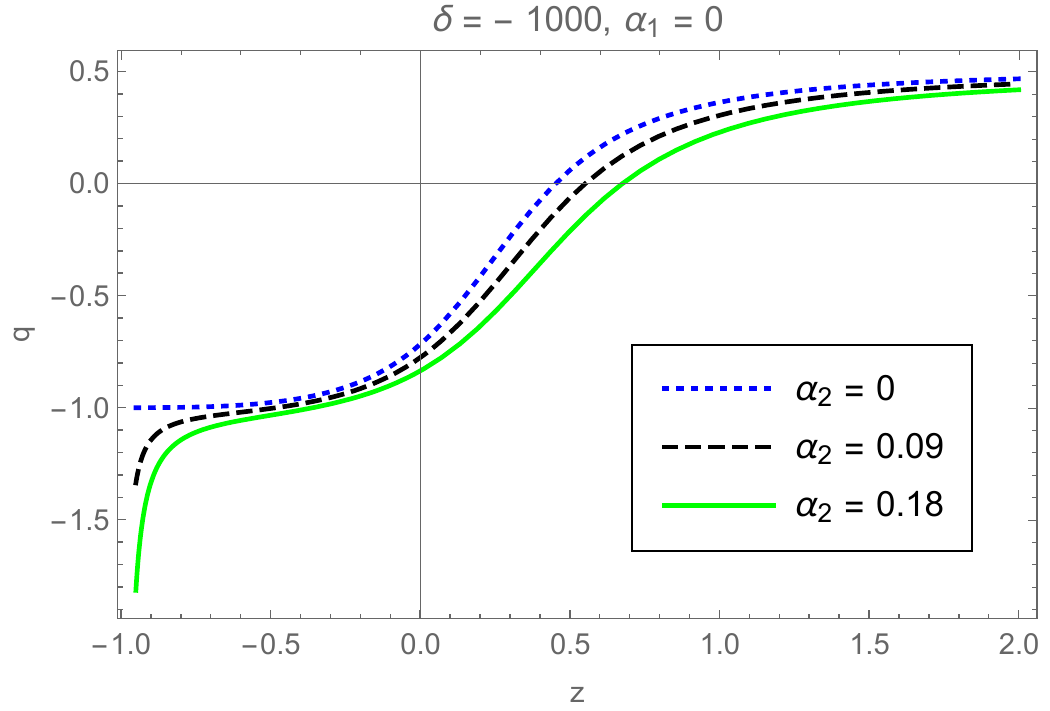}
			(b)\includegraphics[width=5cm, height=5cm]{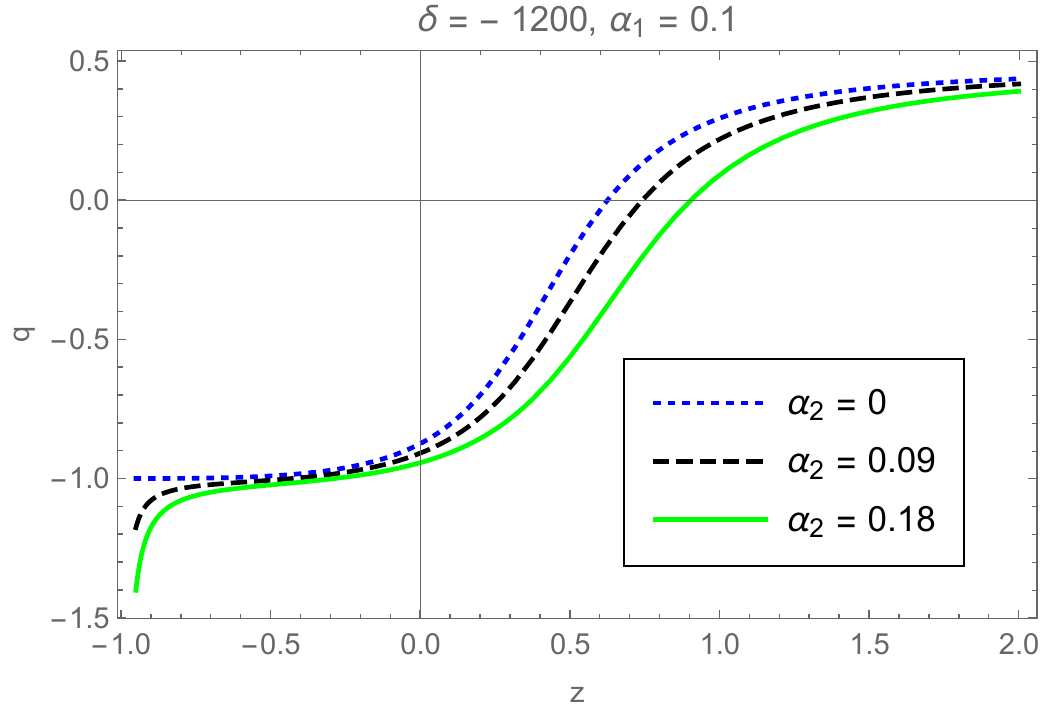}
			(c)\includegraphics[width=5cm, height=5cm]{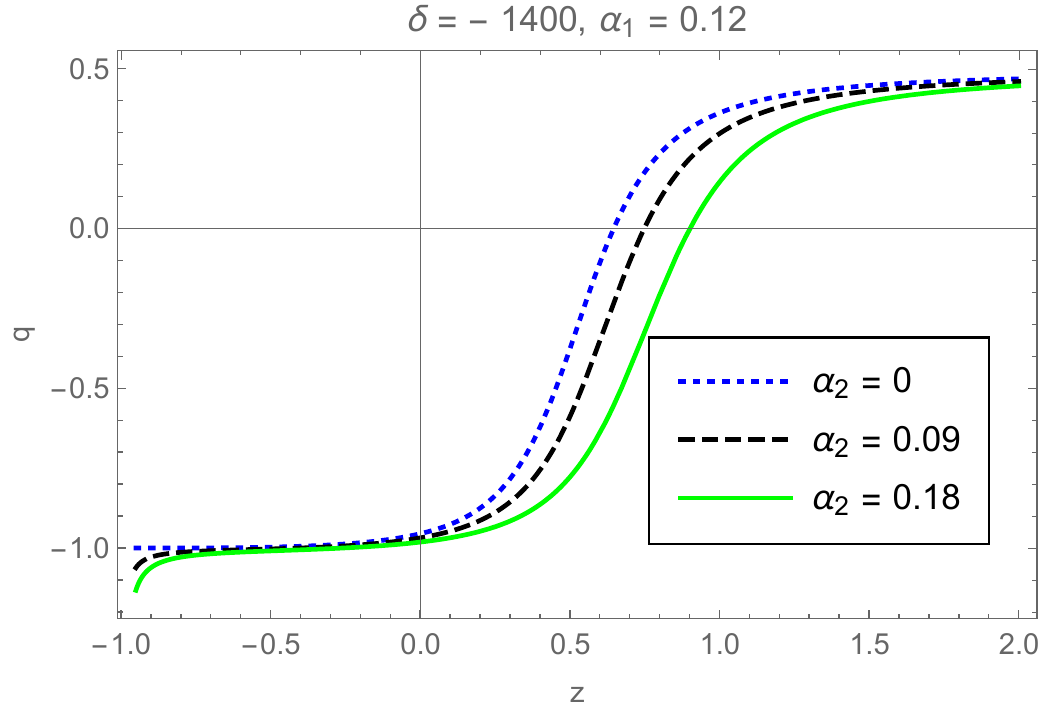}
			
			\caption { The evolution of deceleration parameter ($q$)  in the RHDE model (III) versus redshift $z$ for different values of model parameter $\delta$, $\alpha _1$, and $\alpha _2$ where $H_0$ = 69.6, $\Omega _{D0}$ = 0.70.}

		\end{center}
	\end{figure}

	 For analysis of the RHDE models, we have taken  different values of the parameters $\alpha_1$, $\alpha_2$, and $\delta$. 
		Figs. 1(a), 1(b) and 1(c), depict the behaviour of the deceleration parameter $q$ versus redshift $z$ for different  values of $\alpha_1$, $\alpha_2$, and $\delta$. In Fig. 1(a), we fix the parameters $\alpha_1=0$, and $\delta=-1000$, while we vary $\alpha_2$. It is important to mention that the scenario, $\alpha_1=0$,  $\alpha_2=0$ in this figure and all remaining figures corresponds to the non-interacting case. And, in Figs. 1(b) and 1(c), we vary $\alpha_1$, and $\delta$ for the different  values of $\alpha_2$ as in Fig. 1(a).  We observe from Fig. 1, that $q$ changes its sign from positive to negative. Hence model I shows a transition from  early decelerated phase to present accelerating phase of the Universe for different values of $\alpha_1$, $\alpha_2$  and $\delta$. Fig. 2 shows the evolution of  the EoS parameter $\omega_{D} $ versus redshift $z$ for the different choices of $\alpha_1$, $\alpha_2$, and $\delta$. From Fig. 2(a), we observe that EoS parameter $\omega_{D} $ varies from quintessence to the phantom era $\omega_{D} <-1$ for ($\alpha_1=0$,  $\alpha_2=0$), ($\alpha_1=0$,  $\alpha_2=0.09$) and ($\alpha_1=0$,  $\alpha_2=0.18$).  Finally, converges to the cosmological constant $\omega_{D} = -1$ at future for all the choices of $\alpha_1$, $\alpha_2$. The behaviour for the EoS parameter in Figs. 2(b) and 2(c) for the different choices of $\alpha_1$, $\alpha_2$, and $\delta$ is noticed, which is  different from Fig. 2(a) for $\alpha_2=0.18$. It is also observed from Figs. 2(b) and 2(c), the EoS parameter remains in the phantom era throughout the evolution for  $\alpha_2=0.18$. This indicates that for the interacting RHDE model, $\alpha$ has the linear dependence on $z$. In particular, in \cite{ref55j} it was shown that an interaction is mildly favoured by the combined analysis of several observational data and the equation of state $\omega_{D} $
		could be of phantom character, that means
		$\omega_{D} <-1$.

		  Fig. 3(a), 3(b) and 3(c) describe the behaviour of dark energy density parameter $\Omega _D$ with redshift $z$ for the different choices of $\alpha_1$, $\alpha_2$, and $\delta$. We observe that the Universe is dominated by dark energy at present and also in future. It can also be observed from Fig. 3,  the effect of interaction increases as we increase the value of $\alpha_1$ and decrease the value of $\delta$. 
		 
		  Now, we discuss the stability of the RHDE model. For this purpose, an important quantity is
		  the squared speed of sound $v_s^2$ as we mentioned before.  The $v_s^2\geq0$ (real value of speed), shows a regular propagating mode
		  for a density perturbation. For $v_s^2<0$, the perturbation becomes an irregular wave equation. Hence the negative squared speed (imaginary
		  value of speed) shows an exponentially growing mode for
		  a density perturbation. That is, an increasing density perturbation induces a lowering pressure, supporting the emergence of
		  instability \cite{ref55k,ref55l,ref55m}. The squared speed of the sound $v_s^2$ of the  model I has also been characterized for the stability of the RHDE model.
		 It is plotted in Fig. 4 versus redshift $z$ for the different values of $\alpha_1$, $\alpha_2$, and $\delta$.

		 	From Fig. 4(a), we observe that the RHDE model is not stable for $\alpha_1 = 0$ and $\alpha_2 = 0$ since the velocity becomes imaginary. While the RHDE model is stable between $-0.3<z<0.3$ for $\alpha_1=0$, $\alpha_2=0.09$ and $\alpha_2=0.18$. The velocity becomes imaginary at $z = -0.3$ and $z = 0.3$ for $\alpha_1=0$, $\alpha_2=0.09$ and $\alpha_2=0.18$, at both points, the RHDE model is unstable. Now from Fig. 4(b), it is clear that the RHDE model is stable for all values of $z$ for $\alpha_2=0.09$ and $\alpha_2=0.18$, but for $\alpha_2=0$ the RHDE model is stable only at a late time. Now Fig. 4(c) shows that the RHDE model is stable for all values of $\alpha_1$ and $\alpha_2$ in the region $z<0$. While in the region $z>0$ the velocity becomes imaginary for all values of $\alpha_1$ and $\alpha_2$.  We can also observe from Fig. 4(a),  the squared sound speed $v_s^2$ changes suddenly from close to being $0$ to $4$ at $z=-0.35$, and similarly around $z=0$ in Fig. 4(c). The net effect of increasing $v_s^2$ is higher ISW (Integrated Sachs-Wolfe)
		 	power. This reflects the increased potential decay due
		 	to dark energy; while dark energy perturbations would
		 	help preserve the potential, increasing  $v_s^2$ reduces
		 	the dark energy perturbation contribution and so eases
		 	the decay of the potential. The sqaured sound speed $v_s^2$ is also obtained in terms of the redshift parameter filled with variable modified Chaplygin gas
		 	(VMCG) in  the FRW
		 	Universe  \cite{ref61}. The authors have shown the dependence
		 	of  the velocity of sound $v_s^2$ on the free
		 	parameter $n$ as a function of redshift $z$ \cite{ref61}. In \cite{ref61}, it is observed that the velocity has a magnitude
		 	lying below unity, and
		 	then as the redshift z became smaller, the velocity of
		 	sound increased rapidly for $n < 0$ (phantom dominated
		 	universe). After that the velocity becomes imaginary. They proposed that for $n < 0$ indicates a perturbative cosmology and favors structure formation in the Universe \cite{ref61,ref62}.

	\subsection{ Model II}

	 In model II, the deceleration parameter ($q$) is plotted as a  function of $z$ in Figs. 5(a), 5(b) and 5(c) by considering  different values of the parameter $\alpha _1$, $\alpha _2$ and $\delta $. It also shows that $q$ goes towards  negative from the positive region which depicts the transition of the Universe from early decelerated phase to present accelerating phase for all choices of $\alpha _1$, $\alpha _2$ and $\delta $  similar to the Model I.

	   Figs. 6(a), 6(b) and 6(c) show the evolution of  the EoS parameter $\omega_{D} $ versus redshift $z$ for different values of the parameter $\alpha _1$, $\alpha _2$ and $\delta $. This depicts that the EoS parameter $\omega_{D} $ varies in both quintessence era $\omega_{D} >-1$ and the phantom era $\omega_{D} <-1$. It is also observed that the Model II lies in quintessence era $\omega_{D} >-1$ in the future for some choices of $\alpha _1$, $\alpha _2$ and $\delta $ and in  phantom era $\omega < -1$ for   $\alpha _2=0$  at future. From Figs. 7(a), 7(b) and 7(c), we observe that behaviour of the energy density parameter  $\Omega _D$ in RHDE model (II) versus redshift $z$ for different values of model parameter $\delta$ in  flat FRW Universe for different $\alpha _1$, $\alpha _2$ is different from model I and increases in low redshift region. In Fig. 7, the energy density parameter $\Omega_{D}$ grows larger
	     than 1 for some choices of $\alpha_{2}$, which seems to be
	      impossible according to the definition of $\Omega_{m}$.   This can be  understood as, for  $\rho _D=\frac{3 c^2 H^2}{8 \pi  \left(\frac{\pi  \delta }{H^2}+1\right)}$, the  energy density ratio $r$ is variable, which makes the energy density parameter of the RHDE $\Omega_{D}$ time-dependent. Now from $\omega_{eff}= \Omega_{D}\omega_{D}$ and $	\Gamma=3H(z)\alpha(z)$ relationship, we can say that the increasing value of $\alpha _2$ leads the range of  $\Omega_{D}$ to widen by reason of advance of the dark matter into dark energy.  The  similar conclusion has also been drawn in \cite{ref63,ref64}. From Fig. 8(a), we observe that the RHDE model is unstable $v_s^2 <0$ at present  for all values of $\alpha_1$, $\alpha_2$, and $\delta$.   It can be seen that the RHDE model is stable $v_s^2 \geq0$ in the future  for all values of $\alpha_1$, $\alpha_2$, and $\delta$ as shown in Figs. 8(b) and 8(c).
	   
	   
	   
	   
	     
	
	\begin{figure}
		\begin{center}
			(a)\includegraphics[width=5cm, height=5cm]{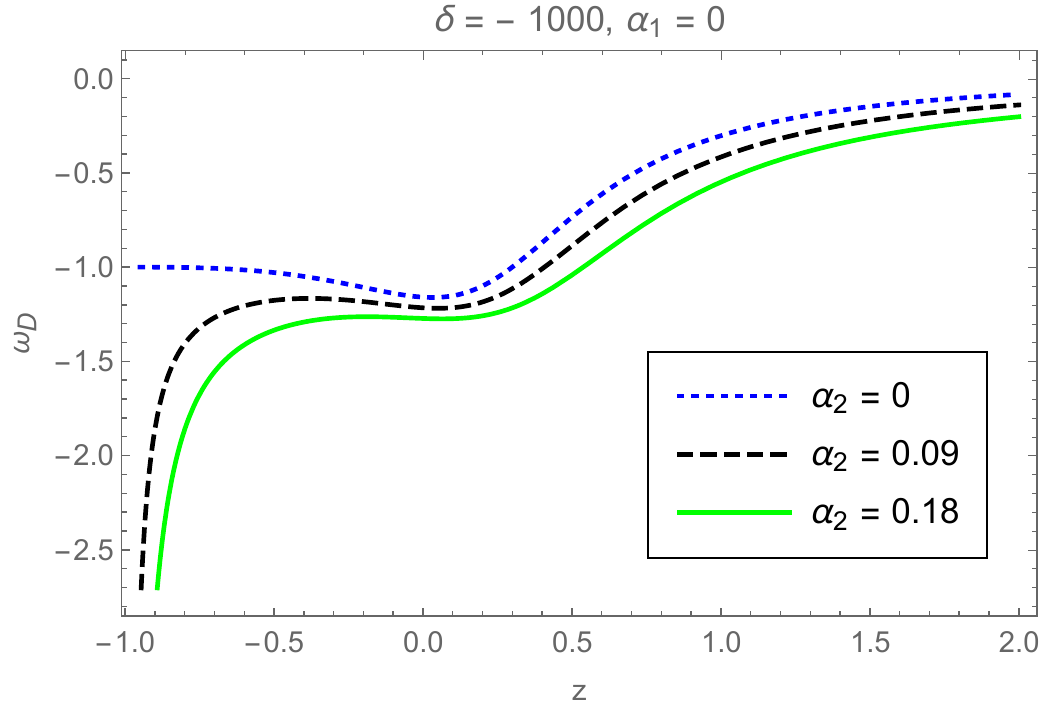}
			(b)\includegraphics[width=5cm, height=5cm]{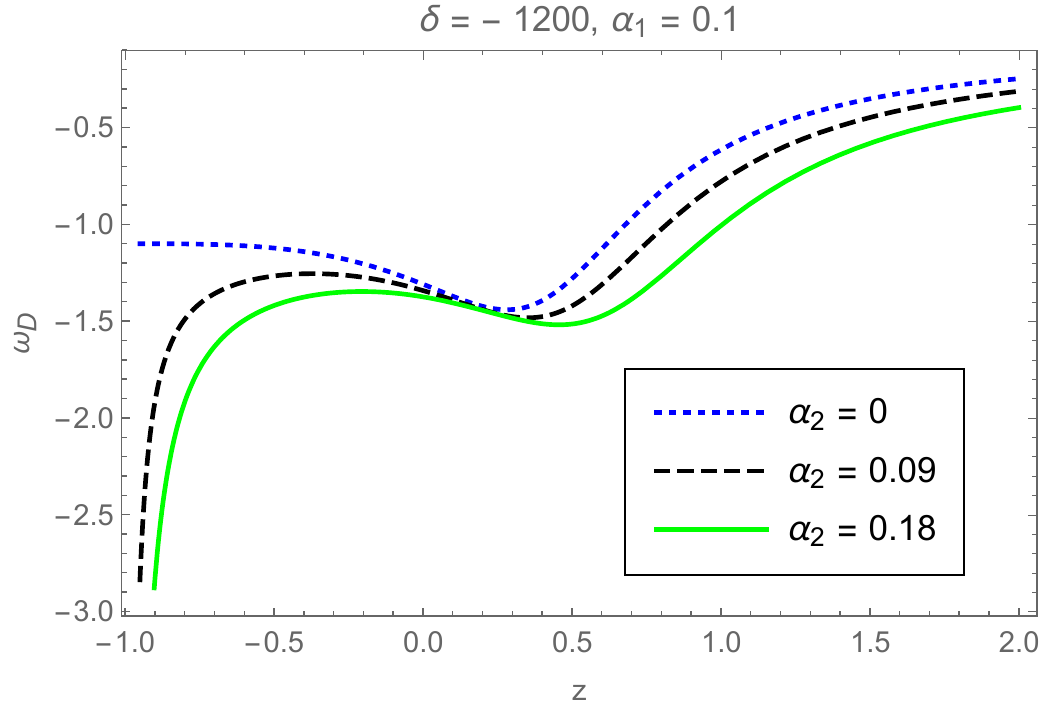}
			(c)\includegraphics[width=5cm, height=5cm]{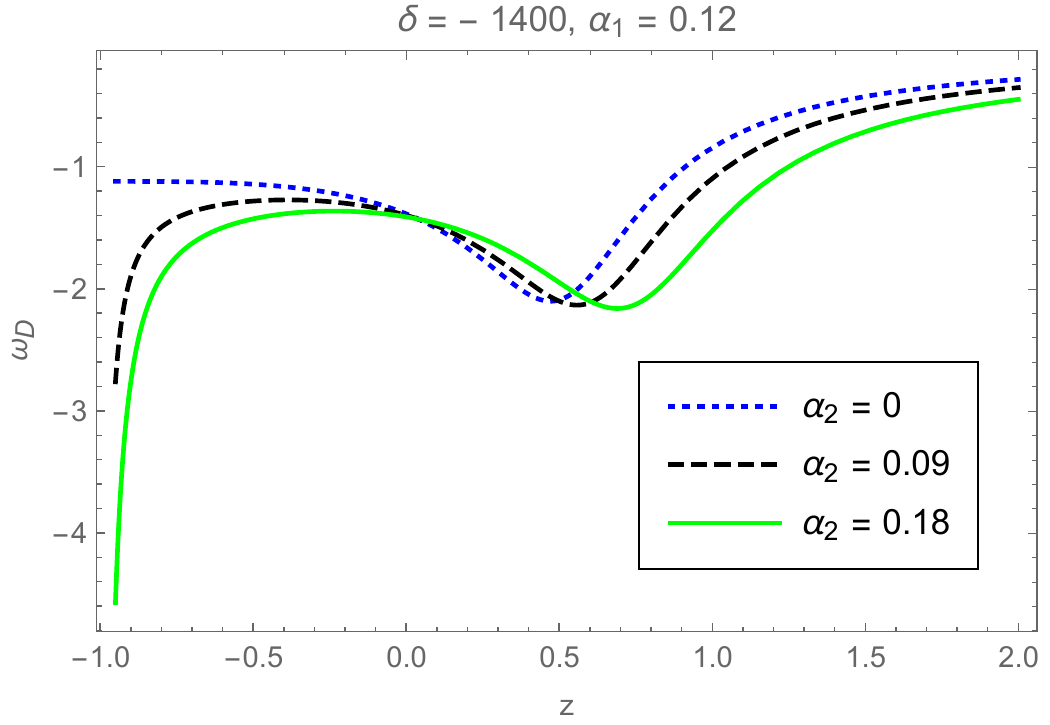}
			
			\caption { The evolution of EOS parameter $\omega _D$ in RHDE model (III) versus redshift $z$ for different values of model parameter $\delta$, $\alpha _1$, and $\alpha _2$ where $H_0$ = 69.6, $\Omega _{D0}$ = 0.70.}

		\end{center}
	\end{figure}
	
	\begin{figure}
		\begin{center}
			(a)\includegraphics[width=5cm, height=5cm]{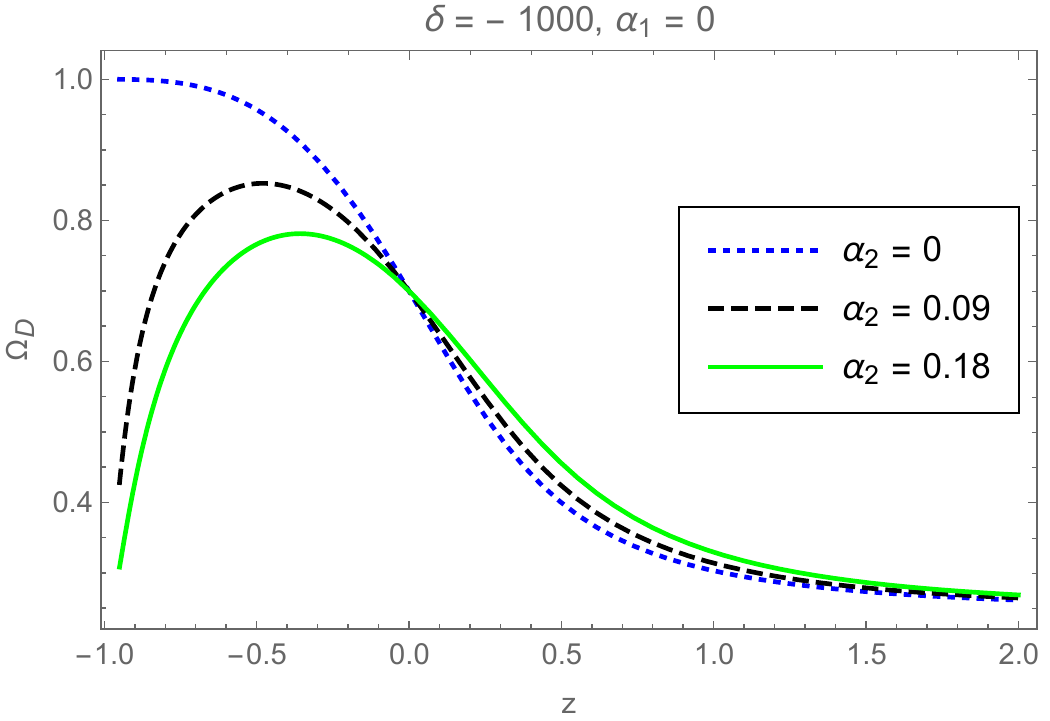}
			(b)\includegraphics[width=5cm, height=5cm]{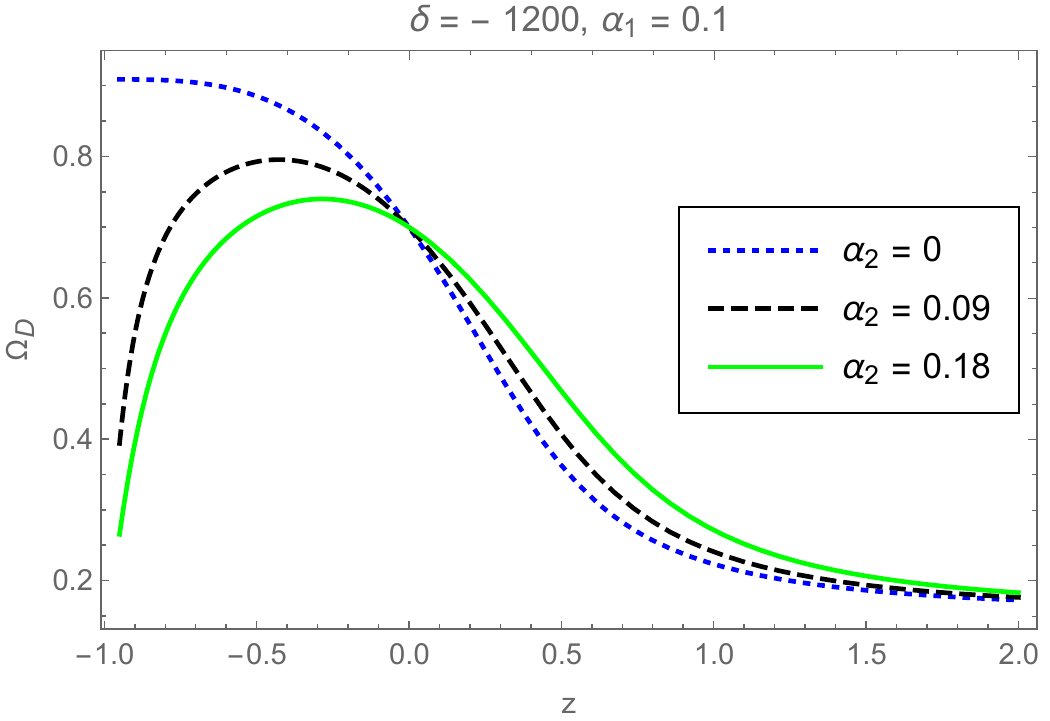}
			(c)\includegraphics[width=5cm, height=5cm]{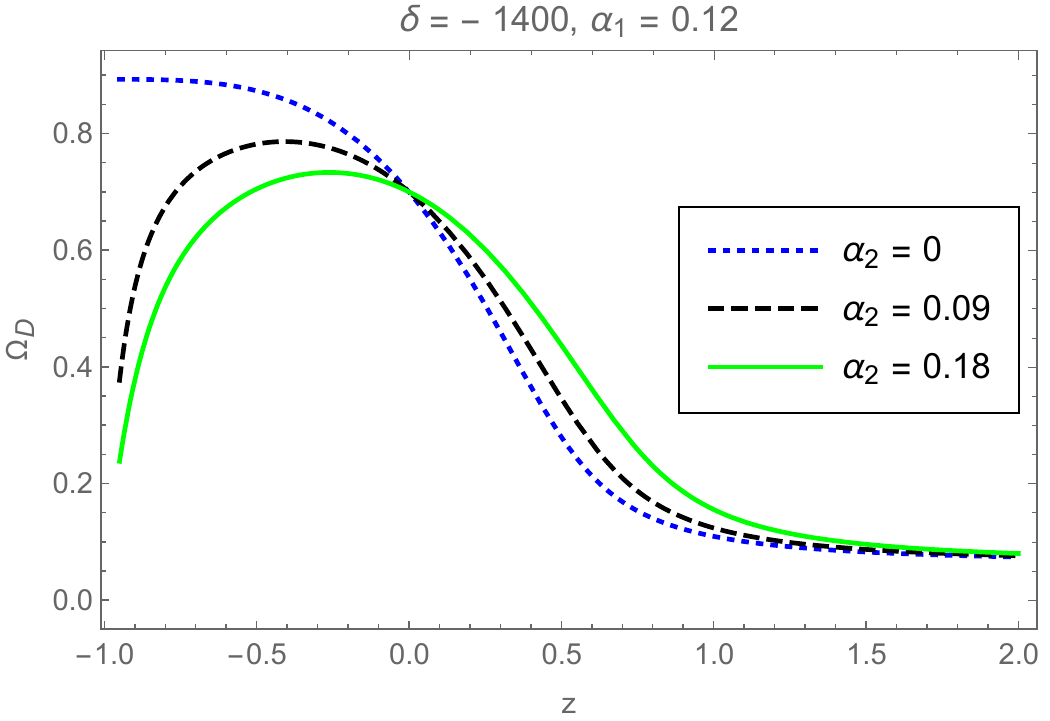}
			
			\caption { The evolution of energy density parameter  $\Omega _D$ in RHDE model (III) versus redshift $z$ for different values of model parameter $\delta$, $\alpha _1$, and $\alpha _2$ where $H_0$ = 69.6, $\Omega _{D0}$ = 0.70.}
			
		\end{center}
	\end{figure}
	
	\begin{figure}
		\begin{center}
			(a)\includegraphics[width=5cm, height=5cm]{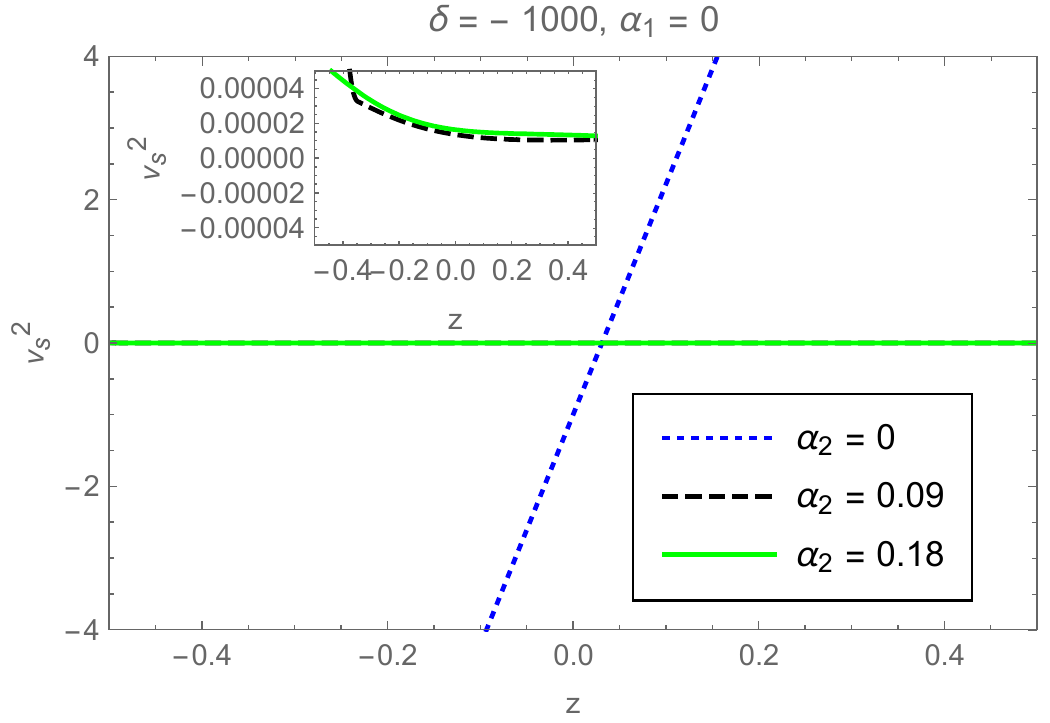}
			(b)\includegraphics[width=5cm, height=5cm]{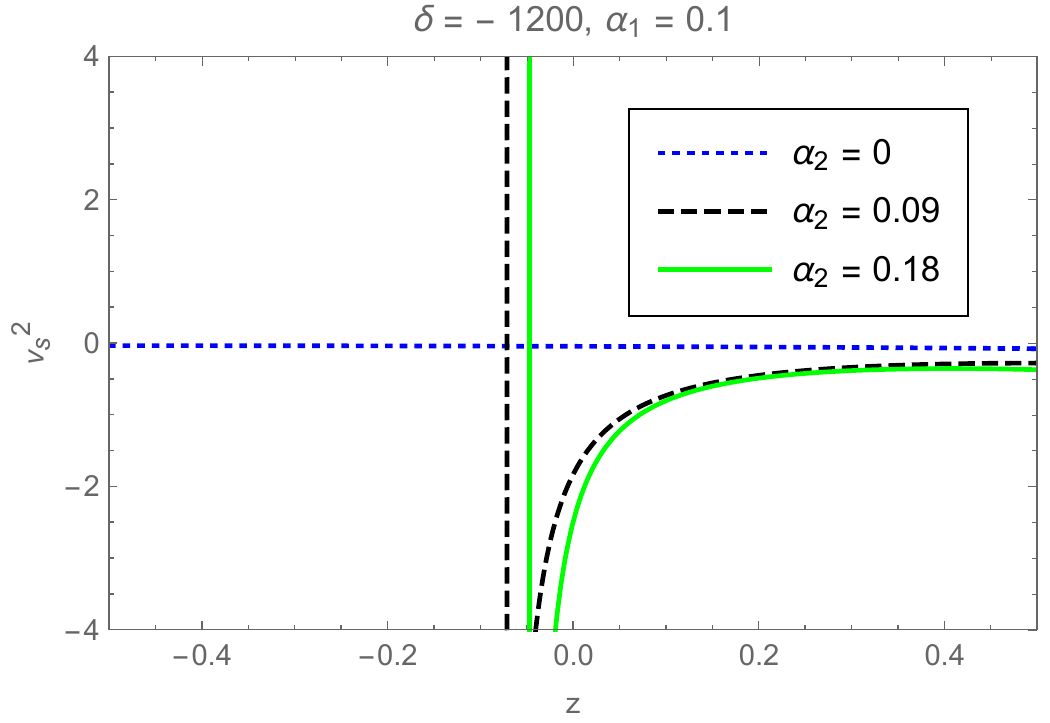}
			(c)\includegraphics[width=5cm, height=5cm]{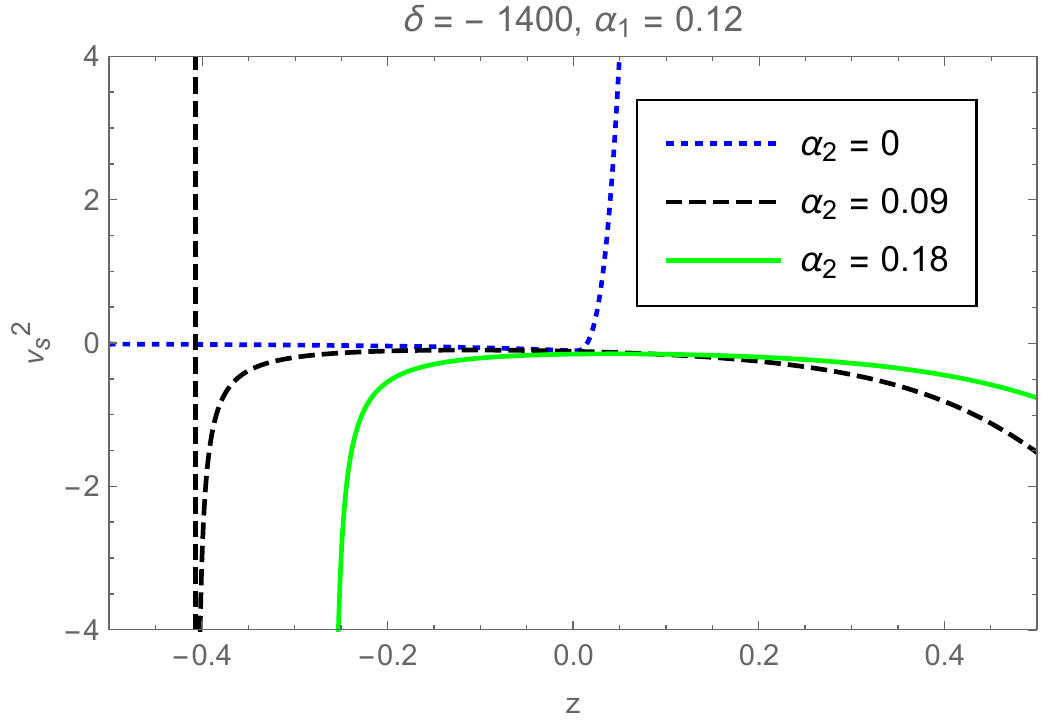}
			
			\caption { The evolution of square of the sound speed parameter $v_s^2$ in RHDE model (III) versus redshift $z$ for different values of model parameter $\delta$, $\alpha _1$, and $\alpha _2$ where $H_0$ = 69.6, $\Omega _{D0}$ = 0.70.}

		\end{center}
	\end{figure}


	\subsection{ Model III}  
		Figs. 9(a), 9(b) and 9(c), depict the behaviour of the deceleration parameter $q$ versus redshift $z$ for different  values of $\alpha_1$, $\alpha_2$, and $\delta$.  We observe from Fig. 9, that $q$ changes its sign from positive to negative for all  values of $\alpha_1$, $\alpha_2$, and $\delta$. Therefore, model III shows a transition from  early decelerated phase to present accelerating phase of the Universe for different  values of $\alpha_1$, $\alpha_2$, and $\delta$. Fig. 10(a), 10(b) and 10(c), exhibit the evolution of  the EoS parameter $\omega_{D} $ versus redshift $z$ for the different choices of $\alpha_1$, $\alpha_2$, and $\delta$. From Fig. 10(a), we observe that EoS parameter $\omega_{D} $ varies from quintessence to the phantom era $\omega_{D} <-1$ for  for all  values of $\alpha_1$, $\alpha_2$, and $\delta$, and remains in the phantom era at future except for ($\alpha_1=0$,  $\alpha_2=0$) for $\delta=-1000$. The EoS parameter remains in the phantom era for all choices of $\alpha_1$, $\alpha_2$, and $\delta$ at future, which can be observed in Figs. 10(b) and 10(c).
		
		Fig. 11(a), 11(b) and 11(c) show the behaviour of dark energy density parameter $\Omega _D$ with redshift $z$ for the different choices of $\alpha_1$, $\alpha_2$, and $\delta$. It can also be observed from Fig. 11, that the variation in both $\alpha_1$ and  $\alpha_2$ contributes to the evolutionary behaviour of the RHDE density parameter $\Omega _D$. This is  particularly noticeable in Fig. 11, the density parameter $\Omega _D$ decreases in the future. In this process, the interaction impacts on both the RHDE and the DM. Accordingly,  their contents are changing due to the energy transfer from the RHDE to the DM. The squared speed of the sound $v_s^2$ versus $z$ of the  model III has been plotted in Fig. 12. From Fig. 12(a), we observe that the RHDE model is unstable $v_s^2 <0$ in the future without interaction i.e.  $\alpha_1=0$ and $\alpha_2=0$.  Also, for $\alpha_2=0.09$  and $\alpha_2=0.18$, the curves look to be overlapping. The inner plot of Fig. 12(a)
		shows a close-up of the outer plot in which the difference can be seen. They are not exactly identical but difference is very small. We can also observe from Fig. 12(b),  that the RHDE model is unstable $v_s^2 \leq0$ in the past  for  $\alpha_2=0.09$, $\alpha_2 =0.18$.  And, the model is stable for $\alpha_2=0$ as can be seen in Fig. 12(c).

	\section{Conclusion}
	This work comprises the study of  the RHDE model where Hubble horizon is taken as the infrared cut-off by taking three different parametrizations of the interaction term in the context of flat FRW Universe.  Different values of the parameters $\alpha_{1}$,  $\alpha_{2}$ and  $\delta$ are taken for the interaction of dark matter and dark energy in the derived model. Following are results which we obtained on the basis of the graphical analysis:\\

		$\ast$
		The sign of deceleration parameter $q(z)$ indicates whether the model inflates or not. The deceleration parameter $q$ shows a transition from early decelerated phase to present accelerating phase of the Universe for all choices of the parameters $\alpha_{1}$, $\alpha_{2}$ and $\delta$ in all three models (I, II, III). The
		deceleration parameter $q(z)$ shows that it increases with redshift and there
		is a transition in the signature of $q(z)$ with the variation in $\alpha_{1}$, $\alpha_{2}$ and $\delta$. \\
		
		$\ast$
		The RHDE equation of state parameter (EoS) $\omega_{D}$ eventually approaches to values negative
		enough to generate the accelerated expansion. The present value of  $\omega_{D}$ remains in phantom regime
		($\omega_{D}\leq-1$) for Model I, Model II and Model III, though the parametrizations are different in these models.  The EoS parameter of   Model I and Model II lies in the phantom region at future, while for Model II remains in quintessence era except  $\alpha_{2}=0$. Therefore,  the equation of state parameter for all three models of the RHDE model shows the different  behaviour in the low-redshift region (at future). \\
		
		$\ast$
		We observe that the dark energy density parameter $\Omega _D$ is a monotonically increasing function of $z$. For the non-interacting case of $\alpha_{1}= \alpha_{2}=0$, we find that $\Omega _D \rightarrow 1$ as $z$ increases as can be seen for all three models,  while  for  the  interaction  case  of $\alpha_{1}= \alpha_{2}\neq0$, $ \Omega _D \neq 1$ for model I,  II and III. The first case is obvious because the R$\acute{e}$nyi holographic energy with the Hubble horizon dominates in the future. Further, the latter shows that two components become comparable, due to the interaction. Similar results have also been found by Kim et al.\cite{ref54a} for the interacting holographic dark energy model with the future event horizon.  We also observe that the Model II is not consistent with the observed evolution scenario of the energy density parameter $\Omega _D$ for $\alpha_{2}=0.09$ and $\alpha_{2}=0.18$. On the other hand model I and III are consistent  with the observed behaviour of the  energy density parameter $\Omega _D$.   \\

		$\ast$
		The graphical behaviour of the squared speed of sound is used to analyse the stability of model I, model II and model III of RHDE.
		We have noticed that the R$\acute{e}$nyi HDE models I, II and III with Hubble horizon as IR cut-off are classically stable ($v_s^2 \geq0$)  for some choices of the parameters $\alpha_{1}$, $\alpha_{2}$ and $\delta$ at present and future. But for some choices of the parameters $\alpha_{1}$, $\alpha_{2}$ and $\delta$, all three models  of the RHDE are unstable $v_s^2 <0$.\\
		
		 In summary, the evolutionary behaviour of the deceleration parameter $(q)$, equation of state parameter EoS ($\omega_{D}$), total energy density parameter ($\Omega_{D}$) and squared sound speed ($v_s^2$) has been investigated for the different choices of $\alpha _1$, $\alpha _2$ and $\delta $. It is observed that the behaviour of the deceleration parameter $(q)$, total energy density parameter ($\Omega_{D}$), the EoS parameter ($\omega_{D}$) and squared sound speed ($v_s^2$) for all three models are different from high red-shift region to low redshift region for some choices of $\alpha _1$, $\alpha _2$ and $\delta $. While, the behaviour of cosmological parameters of model III is similar  to the  model I and model II for some choices of $\alpha _1$, $\alpha _2$ and $\delta $. \\

		For the future work, different observational data sets such as  the distance modulus measurements of type Ia supernova from the Joint Light-curve Analysis  and the observational measurements of Hubble parameter  can be utilized  to reconstruct the model  parameters to figure out the correct model and to understand the nature of the RHDE.

	\section*{Acknowledgements}
	
	 We are very much grateful to the honourable referee for
	the precious time and illuminating comments that have significantly
	improved our work in terms of research quality and
	presentation.

\end{document}